\documentclass[10pt,aps,prx,twocolumn,floats,epsfig,showpacs]{revtex4-1}

\usepackage{amsmath}
\usepackage{amsfonts}
\usepackage{exscale}
\usepackage{float}
\usepackage{tikz}

\usepackage{bbm}
\usepackage{amssymb}
\usepackage{ifpdf}
\usepackage{color}
\usepackage{graphicx,epsfig}

\newcommand{\N}{{\mathbb{N}}}
\newcommand{\Z}{{\mathbb{Z}}}
\newcommand{\R}{{\mathbb{R}}}

\newcommand{\1}{{\mathbbm{1}}}
\newcommand{\p}{\partial}

\begin{document}

\title{Canonical quantization on the half-line and in an interval based upon \\
a new concept for the momentum in a space with boundaries}

\author{M.\ H.\ Al-Hashimi and U.-J.\ Wiese$^\dagger$}
\affiliation{$^\dagger$ Albert Einstein Center, Institute for Theoretical 
Physics, University of Bern, 3012 Bern, Switzerland}

\begin{abstract}
For a particle moving on a half-line or in an interval the operator 
$\hat p = - i \p_x$ is not self-adjoint and thus does not qualify as the 
physical momentum. Consequently canonical quantization based on $\hat p$ fails. 
Based upon a new concept for a self-adjoint momentum operator $\hat p_R$, 
we show that canonical quantization can indeed be implemented on the half-line 
and on an interval. Both the  Hamiltonian $\hat H$ and the momentum operator 
$\hat p_R$ are endowed with self-adjoint extension parameters that characterize 
the corresponding domains $D(\hat H)$ and $D(\hat p_R)$ in the Hilbert space. 
When one replaces Poisson brackets by commutators, one obtains meaningful 
results only if the corresponding operator domains are properly taken into 
account. The new concept for the momentum is used to describe the results of 
momentum measurements of a quantum mechanical particle that is reflected at
impenetrable boundaries, either at the end of the half-line or at the two ends 
of an interval.

\end{abstract}

\maketitle

\section{Introduction}

Momentum is one of the most fundamental physical quantities. The momentum 
operator generates infinitesimal translations in infinite space. When the
Hamiltonian is translation invariant, momentum is conserved. Coordinates and
their canonically conjugate momenta play a central role in Hamiltonian dynamics,
which turns into quantum mechanics upon canonical quantization. 

Several important physical systems are confined inside a limited region of space
with sharp boundaries. These include, for example, ultra-cold atoms in an 
optical box trap \cite{Gau13} of a few tens of micrometers in size, electrons in
a quantum dot at the nanometer scale \cite{Har05}, the ``femto-universe'' of the
phenomenological MIT bag model \cite{Cho74,Cho74a,Has78} for confined quarks 
and gluons, domain wall fermions \cite{Kap92,Sha93}, or intervals of 
extra-dimensional space at the Planck scale. In the idealized mathematical 
description of such systems, space is endowed with sharp impenetrable 
boundaries. While for most systems this is just a mathematical convenience 
that allows one to exclude regions of very high potential energy, it is 
conceivable that extra-dimensional space literally ends at a boundary. In 
spaces with boundaries, translation invariance is explicitly broken, not just 
because the Hamiltonian includes symmetry-breaking terms, but because space 
itself ends (at least in the idealized mathematical description). 
As a consequence, the usual quantum mechanical momentum operator 
$\hat p = - i \p_x$ is no longer self-adjoint, and thus no longer represents 
the physical momentum. From this it has been concluded that, in a finite 
volume, momentum is no longer a measurable physical observable \cite{Bon01}. As 
another consequence of the non-self-adjointness of $\hat p$, canonical 
quantization (which is based upon $\hat p$) fails for confined systems with 
sharp boundaries. 

Recently, we have introduced a new concept for the momentum of a quantum
mechanical particle in a box \cite{AlH20}, which gives rise to a physically
and mathematically satisfactory self-adjoint momentum operator. This concept
naturally extends to the half-line as well as to higher dimensions. Intervals 
and half-lines are the basic building blocks of higher-dimensional regions of
space with sharp boundaries. The present investigations hence form the basis
for future applications to physical systems in higher dimensions. The sharp
boundaries of a confined system give rise to a high degree of ultraviolet 
sensitivity. At low energies, this reflects itself in the values of self-adjoint
extension parameters of the Hamiltonian that characterize the boundary 
conditions of the wave function. 

In order to take the ultraviolet sensitivity 
into account properly, in our construction of a self-adjoint momentum operator 
in \cite{AlH20} we started out from an ultraviolet lattice regularization. 
While this is familiar in non-perturbative quantum field theory, in this case 
it is beneficial even in quantum mechanics. On the lattice, one naturally 
distinguishes forward and backward derivatives, neither of them being 
Hermitean. The Hermitean lattice momentum operator is given by a symmetrized 
forward-backward derivative, which manifests itself as a finite difference that 
extends over two lattice spacings. This naturally leads to the distinction of 
even and odd lattice points. A careful analysis of the problem reveals that the 
construction of a self-adjoint momentum operator in the continuum limit 
requires the doubling of the standard Hilbert space, in order to maintain a 
remnant of the crucial distinction between even and odd lattice points even in 
the continuum \cite{AlH20}. In fact, some aspects of the problem are 
reminiscent of the lattice fermion doubling problem \cite{Nie81,Nie81a,Fri82}, 
which arises because the Dirac operator contains first-order derivatives. The 
resulting insight of \cite{AlH20} is that the construction of a self-adjoint 
momentum operator requires a refined concept, not only of Hilbert space, but 
even of space itself. This may not be too surprising, because a space that ends
abruptly supports other momentum modes than infinite space. Still, as we will
see, the finite-energy sector, which is defined by the Hamiltonian, resides in 
a region of Hilbert space that is completely equivalent to the standard quantum 
mechanical treatment.

An extension of the Hilbert space also plays an important role in quantum 
measurements related to a positive operator-valued measure (POVM) 
\cite{Dav70,Kra71,Hol78,Hol79,Dav76,Hol01,Kra83}, which are based on Kraus
operators describing quantum jumps. Measurement processes using a POVM provide
a generalization of von Neumann's standard projective measurements. They play 
an important role for controlling quantum systems and processing quantum 
information. In POVM-based measurements, the quantum system to be investigated 
is first coupled to another quantum system that acts as an ancilla. Then a 
standard projective measurement is performed on the ancilla, which 
indirectly affects the quantum system under study \cite{Per90}. That a given 
POVM can be realized by an appropriate extension of the Hilbert space is 
related to Naimark's theorem \cite{Nai40,Nai43,Akh63,Hel73,Hol01}. In this way 
measurements on confined systems have been described by POVMs \cite{Bel09} and 
an optimal POVM for a particle on a half-line has been considered in 
\cite{Shi08}. It should be pointed out that the POVM measurement is not based 
on a self-adjoint momentum operator of the quantum particle itself, but rather 
of the particle coupled to its ancilla. Our construction, on the other hand, 
provides a self-adjoint momentum operator for the particle alone (without 
invoking any ancilla). In our case, the doubling of the Hilbert space results 
from the necessity to include states whose energy is ultraviolet-sensitive, but 
which, due to the existence of sharp boundaries, still contribute to the 
momentum eigenstates. In other words, the Hilbert space that contains the 
finite-energy eigenstates is too small to contain also the momentum eigenstates 
of a quantum particle in a space with sharp boundaries.

The time-evolution of a quantum system is driven by its Hamiltonian, which is
described by a self-adjoint operator acting in an appropriate Hilbert space. In
non-relativistic 1-d quantum mechanics, the single-particle Hamiltonian
$\hat H = - \tfrac{1}{2m} \p_x^2 + V(x)$ (in units where $\hbar = 1$) contains 
the differential operator $\p_x^2$, which is supposed to act on 
square-integrable wave functions. Since not all square-integrable functions
are differentiable, the Hamiltonian (as well as other physical operators) act 
only in a restricted domain $D(\hat H)$ of the Hilbert space. For a differential
operator the domain is characterized by the square-integrability of the 
corresponding derivatives of the wave function. In an infinite-dimensional 
Hilbert space, there are subtle differences between Hermiticity and 
self-adjointness, which were first understood by von Neumann \cite{Neu32a}. 
Hermiticity means that an operator $\hat A$ and its adjoint $\hat A^\dagger$ act 
in the same way. Self-adjointness requires, in addition, that the corresponding 
domains $D(\hat A) = D(\hat A^\dagger)$ coincide \cite{Ree75,Gie00,Jur21}. In 
order to qualify as a physical observable, an operator must be self-adjoint. 
This is because only self-adjointness, and not Hermiticity alone, guarantees a 
spectrum of real eigenvalues with a corresponding complete set of orthonormal 
eigenfunctions. During its time-evolution the wave function of a particle with 
finite energy only explores the domain $D(\hat H)$, and never reaches other 
corners in Hilbert space.

However, when the unitary time-evolution driven by the Hamiltonian is 
interrupted by an (idealized) projective measurement, the momentary wave 
function is projected onto an eigenstate of the operator $\hat A$ that describes
the measured observable. It is possible that the domains of the Hamiltonian 
$D(\hat H)$ and of the observable $D(\hat A)$ do not coincide. This is no 
problem, because a self-adjoint operator has a complete set of eigenfunctions. 
As a consequence, every wave function (even outside of $D(\hat A)$) can be 
represented arbitrarily well by a superposition of eigenstates of $\hat A$. In 
this way, one can determine the probabilities to measure the various possible 
eigenvalues of the observable $\hat A$. After such a projective measurement, 
the wave function is inside $D(\hat A)$ but not necessarily any longer inside 
$D(\hat H)$. How can the unitary time-evolution proceed after such a 
measurement? Again, since the Hamiltonian is self-adjoint, any state (even 
outside $D(\hat H)$) can be approximated arbitrarily well by a superposition of 
eigenstates of $\hat H$, and the time-evolution proceeds accordingly. However, 
the energy expectation value after a measurement that leads out of $D(\hat H)$ 
is usually infinite. Hence, idealized measurements can transfer an infinite 
amount of energy to the particle under investigation. Of course, any practical 
measurement only consumes a finite amount of energy, and is, in any case, not 
completely realistically described by an idealized projective measurement.

Momentum measurements on confined particles fall in this category. For example,
a particle of finite energy that moves along the entire real axis in a 
potential that diverges at spatial infinity, $V(\pm \infty) \rightarrow \infty$,
has a square-integrable wave function that vanishes at infinity and belongs to
the Hilbert space $L^2(\R)$ of square-integrable wave functions over the entire
real axis. When the momentum of the particle is measured and one obtains the 
value $k$, its wave function collapses onto the plane wave momentum eigenstate 
$\langle x|k\rangle = \exp(i k x)$, which has $\langle k|V|k\rangle = \infty$. 
Hence, an idealized momentum measurement indeed transfers an infinite amount of 
energy to a confined particle.

The situation becomes more subtle in the presence of impenetrable sharp
boundaries (which, of course, again are mathematical idealizations) 
\cite{Bal70,Cla80,Far90,Car90,Bon01,AlH12}. For example, for a particle that is 
strictly confined to the positive real axis (the half-line) the operator 
$\hat p = - i \p_x$ is not self-adjoint, and hence it has until now been 
unknown how to properly define the corresponding momentum operator. As a 
consequence, one has replaced the momentum operator by the dilation operator, 
thus moving from canonical to affine quantization 
\cite{Twa06,Kla12,Ber14,Alm18,Gou20}. The main purpose of this paper is to 
provide an appropriate construction of a self-adjoint momentum operator that is 
satisfactory both from a physical and from a mathematical point of view, and to 
show that canonical quantization is, in fact, applicable to the half-line as 
well as to an interval. As a result, we will be able to describe momentum 
measurements performed on a quantum mechanical particle that is strictly 
limited to the positive real axis or to an interval, even after a momentum 
measurement. Such a particle is bound to or reflected at impenetrable 
boundaries, either at the origin or at the two ends of the interval.

It should be pointed out that our construction of a self-adjoint momentum 
operator assumes that some physically meaningful ultraviolet cut-off (like a 
crystal lattice in a quantum dot) actually exists, and that the quantum 
mechanical description in the continuum is an effective low-energy description 
that is valid only below that ultraviolet cut-off. If one assumes that the 
quantum mechanics formulated in the continuum is a ``theory of everything'', 
in other words that no physical ultraviolet cut-off exists at short distances, 
there would be no basis for extending the Hilbert space. In that case, momentum 
would indeed not be a meaningful concept for motion on the half-line, and one
would be restricted to affine quantization. In experimental situations 
involving, for example, quantum dots, there is always a physical ultraviolet
cut-off at which the effective low-energy quantum mechanical description in the
continuum breaks down, such that our momentum concept is indeed applicable. 
Whether the Planck length leads to a physical ultraviolet cut-off in 
extra-dimensional spaces with boundaries is, of course, a matter of speculation.

The paper is written also with some pedagogical intentions in mind. Therefore
we do not assume that the reader is familiar with the concept of operator
domains, which is crucial for the distinction between Hermiticity and 
self-adjointness. Unfortunately, in the education of the typical theoretical
physicist these issues often do not play a prominent role. The experts will
hopefully not be offended that we elaborate on some issues that are well-known
to them. We also like to point out that the notion of canonical quantization is
not as uniquely defined as one might think. Definitely, it describes 
quantization in an equal-time Hamiltonian formulation, rather than, for 
example, on a light-cone or some other hyper-surface. Here we use a more narrow 
definition of canonical quantization, which is based upon the canonical 
commutation relations between coordinates and conjugate momenta (or closely
related variants thereof). This definition distinguishes canonical quantization 
from affine quantization, which also operates in an equal-time Hamiltonian 
framework, but replaces the momentum by the dilation operator. The main new 
result of our work is the construction of an appropriate self-adjoint momentum 
operator, which forms the basis for successfully applying canonical 
quantization to the half-line and to an interval, for which it was thought to 
be inapplicable. Still, applying canonical quantization (defined in this way) 
to the half-line or an interval is less straightforward than for the entire 
real axis, because some subtleties related to operator domains are crucial. 
Although, this is well-known to the experts, we will discuss explicitly how 
canonical quantization should be applied in such cases.

The rest of the paper is organized as follows. In Section II, we address the 
non-self-adjointness of the standard momentum operator $\hat p = -i \p_x$ on the
half-line $\R_{\geq 0}$, we construct the self-adjoint extensions of the
Hamiltonian, and we address canonical quantization and standard 
momentum measurements, as well as affine quantization. Based upon our new
concept, in Section III we construct a self-adjoint momentum operator on the 
half-line, embed the Hamiltonian in the resulting mathematical framework, and 
discuss the corresponding momentum measurements. We then consider the resulting 
canonical quantization on the half-line including the classical limit. Section
IV addresses canonical quantization on an interval and relates the results to
the situation on a circle. Finally, we end with some conclusions. Ultraviolet
lattice aspects of the new momentum concept are summarized in an appendix.

\section{From canonical quantization on the entire real
axis to affine quantization on the half-line}

In this section we address the non-self-adjointness of the standard momentum
operator on the half-line and its consequences for canonical and affine
quantization.

\subsection{Non-self-adjointness of $-i \p_x$ on $\R_{\geq 0}$}

Let us consider the standard momentum operator $\hat p = - i \p_x$ on the 
half-line $\R_{\geq 0}$. Using partial integration one obtains
\begin{equation}
\langle \hat p^\dagger \chi|\Psi\rangle = \langle\chi|\hat p \Psi\rangle = 
\langle \hat p \chi|\Psi\rangle - i \chi(0)^* \Psi(0) \ .
\end{equation}
Hermiticity requires that $\chi(0)^* \Psi(0) = 0$. This requirement can be 
satisfied if one restricts the domain $D(\hat p)$ to those wave functions whose 
derivative is square-integrable and that obey $\Psi(0) = 0$. However, then
$\chi(0)$ remains unrestricted and can still assume arbitrary values. 
Consequently, the domain of $\hat p^\dagger$, which acts on $\chi$, remains 
unrestricted, and $D(\hat p^\dagger) \supset D(\hat p)$. When $\Psi(0)$ = 0, 
$\hat p$ and $\hat p^\dagger$ act in the same way and hence $\hat p = - i \p_x$ 
is indeed Hermitean. However, since $D(\hat p^\dagger) \neq D(\hat p)$, it is 
not self-adjoint. In fact, it is impossible to extend $\hat p$ to a self-adjoint
operator on the half-line. Consequently, $\hat p = - i \p_x$ does not describe 
the physical momentum of a quantum mechanical particle that moves along the 
positive real axis. As a result, it has been concluded that, in this case, 
momentum is no longer an observable physical quantity \cite{Bon01}. We will
reach a different conclusion, namely that not $\hat p = - i \p_x$ (which is not 
self-adjoint) but another operator, $\hat p_R = - i \sigma_1 \p_x$, which is 
self-adjoint in the Hilbert space $L^2(\R_{\geq 0}^2)$ of the doubly-covered 
positive real axis, describes the physical momentum of a particle on the 
half-line. In fact, the appropriate momentum operator $\hat p_R + i \hat p_I$ 
has a Hermitean component $\hat p_R$ as well as an anti-Hermitean component 
$i \hat p_I$, with both $\hat p_R$ and $\hat p_I$ being self-adjoint.

\subsection{Self-adjoint extension of $\hat H$ on 
$\R_{\geq 0}$}

When restricted to the half-line, the self-adjointness of 
$\hat H = - \tfrac{1}{2 m} \p_x^2 + V(x)$, with $V(x)$ being non-singular, 
requires the following adaptations. First of all, by performing two partial 
integrations one obtains
\begin{eqnarray}
&&\langle \hat H^\dagger \chi|\Psi\rangle = \langle\chi|\hat H \Psi\rangle = 
\nonumber \\ 
&&\langle \hat H \chi|\Psi\rangle - \frac{1}{2 m} 
\left[\p_x \chi(0)^* \Psi(0) - \chi(0)^* \p_x \Psi(0)\right] \ .
\label{HHermiticity}
\end{eqnarray}
Hermiticity hence requires the term in square-brackets to vanish. The most
general boundary condition that is consistent with the linearity of quantum
mechanics is the Robin boundary condition
\begin{equation}
\gamma \Psi(0) - \p_x \Psi(0) = 0 \ .
\label{Robinbc}
\end{equation}
Dirichlet boundary conditions, $\Psi(0) = 0$, result from 
$\gamma \rightarrow \infty$, while Neumann boundary conditions, 
$\p_x \Psi(0) = 0$, correspond to $\gamma = 0$. Since (for finite $\gamma$) 
$\Psi(0)$ itself can still take arbitrary values, inserting eq.(\ref{Robinbc}) 
in the square-bracket in eq.(\ref{HHermiticity}), the Hermiticity condition 
turns into
\begin{equation}
\left[\p_x \chi(0)^* - \gamma \chi(0)^*\right] \Psi(0) = 0 \ \Rightarrow \
\gamma^* \chi(0) - \p_x \chi(0) = 0 \ .
\end{equation}
This relation characterizes the domain $D(\hat H^\dagger)$ of $\hat H^\dagger$ 
(which acts on $\chi$). The Hamiltonian is self-adjoint only if the two domains 
coincide, $D(\hat H^\dagger) = D(\hat H)$, i.e.\ if $\gamma^* = \gamma \in \R$. 
In this way, we obtain a 1-parameter family of self-adjoint extensions of 
$\hat H$. 

Self-adjointness is not just a mathematical requirement, it also has most
important physical consequences. In particular, for $\gamma \in \R$ the 
boundary condition of eq.(\ref{Robinbc}) ensures that the probability current
density
\begin{equation}
j(x) = \frac{1}{2 m i} [\Psi(x)^* \p_x \Psi(x) - \p_x \Psi(x)^* \Psi(x)] \ ,
\end{equation}
vanishes at the boundary, i.e.\ $j(0) = 0$, and hence does not flow into the
forbidden region on the negative real axis. This ensures unitarity, i.e.\ 
probability conservation, for the particle moving along the half-line.

In the absence of a potential ($V(x) = 0$) it is easy to construct the energy 
eigenstates. First of all, there are stationary scattering states of positive 
energy $E = \tfrac{p^2}{2 m}$
\begin{equation}
\psi_E(x) = \exp(- i p x) + R(p) \exp(i p x) \ , \ 
R(p) = \frac{i p + \gamma}{i p - \gamma}.
\label{Requation}
\end{equation}
It is sufficient to limit oneself to $p \geq 0$, because (together with a bound
state for $\gamma < 0$) these states alone form a complete orthonormalized set 
with 
\begin{equation}
\langle \psi_{E'}|\psi_E\rangle = 2 \pi \delta(p - p') \ ,
\end{equation}
where $E' = \tfrac{{p'}^2}{2m}$, $p' > 0$. In particular, since 
$R(- p) = R(p)^* = R(p)^{-1}$, the states with opposite values of $p$ are simply
given by
\begin{equation}
\exp(i p x) + R(- p) \exp(- i p x) = R(- p) \psi_E(x) \ .
\end{equation}

In addition, for $\gamma < 0$, there is a bound state of negative energy
\begin{equation}
\psi_b(x) = \sqrt{- 2 \gamma} \exp(\gamma x) \ , \quad
E_b = - \frac{\gamma^2}{2 m} \ .
\end{equation}
Interestingly, a perfectly reflecting impenetrable barrier can still support
bound states, and the Hamiltonian $\hat H = - \frac{1}{2 m} \p_x^2$ (endowed 
with a negative self-adjoint extension parameter $\gamma < 0$) indeed has an
eigenstate of negative energy. It is easy to convince oneself that the bound
state is orthogonal to the scattering states, i.e.\ 
$\langle\psi_E|\psi_b\rangle = 0$.

\subsection{Canonical quantization on $\R$}

\begin{figure}[tbp]
\vspace{3cm}
\begin{tikzpicture}
\tikz\draw
[line width=0.4mm,color=black] (0,0) -- (8,0)
node[right] {$\hat x$}
[line width=0.4mm,color=black] (7.9,0.1) -- (8,0)
[line width=0.4mm,color=black] (7.9,-0.1) -- (8,0)
[line width=0.4mm,color=black] (6,-0.2) -- (6,0.2)
node[above] {$|x\rangle$}
[line width=0.4mm,color=black] (3,-0.2) -- (3,0.2)
node[above] {$|x - a\rangle$}
[line width=0.4mm,color=black] (3,-0.5) -- (6,-0.5)
node[below] {\hspace{-3cm} $U_a$}
[line width=0.4mm,color=black] (3.1,-0.4) -- (3,-0.5)
[line width=0.4mm,color=black] (3.1,-0.6) -- (3,-0.5)
[line width=0.4mm,color=black] (0,-2) -- (8,-2)
node[right] {$\hat p$}
[line width=0.4mm,color=black] (7.9,-1.9) -- (8,-2)
[line width=0.4mm,color=black] (7.9,-2.1) -- (8,-2)
[line width=0.4mm,color=black] (5,-2.2) -- (5,-1.8)
node[above] {$|k + q\rangle$}
[line width=0.4mm,color=black] (2.5,-2.2) -- (2.5,-1.8)
node[above] {$|k\rangle$}
[line width=0.4mm,color=black] (2.5,-2.5) -- (5,-2.5)
node[below] {\hspace{-2.5cm} $\widetilde U_q$}
[line width=0.4mm,color=black] (4.9,-2.4) -- (5,-2.5)
[line width=0.4mm,color=black] (4.9,-2.6) -- (5,-2.5);
\end{tikzpicture}
\vspace{-3.5cm}
\caption{\it Action of the translation operators $U_a = \exp(i \hat p a)$ and
$\widetilde U_q = \exp(i q \hat x)$ on the position and momentum eigenstates 
for the entire real axis $\R$.}
\label{Fig1}
\end{figure}
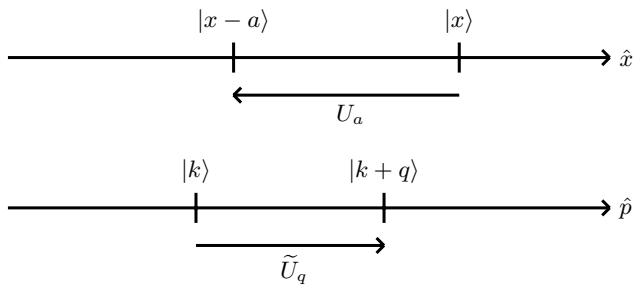

Canonical quantization is based upon the canonical commutation relation 
$[\hat x,\hat p] = i$. This relation applies to unrestricted linear motion, 
because in the Hilbert space $L^2(\R)$ the momentum operator 
$\hat p = - i \p_x$ is indeed self-adjoint. The operator $\hat p$ then 
generates infinitesimal translations in coordinate space. The unitary operator 
that translates a wave function by a distance $- a$ acts as
\begin{equation}
U_a \Psi(x) = \exp(i \hat p a) \Psi(x) = \exp(a \p_x) \Psi(x) = \Psi(x + a) \ .
\label{coordinateshift}
\end{equation}
By a Fourier transformation, we obtain the momentum space wave function
\begin{eqnarray}
\widetilde \Psi(p)&=&\langle p|\Psi\rangle = 
\int_{- \infty}^\infty dx \langle p|x\rangle \langle x|\Psi\rangle \nonumber \\
&=&\int_{- \infty}^\infty dx \exp(- i p x) \Psi(x) \ .
\end{eqnarray}
The position operator $\hat x = i \p_p$ generates infinitesimal translations in 
momentum space. The unitary operator $\widetilde U_q = \exp(i q \hat x)$, which 
translates a momentum space wave function by $q$, acts as
\begin{equation}
\widetilde U_q \widetilde\Psi(p) = \exp(- q \p_p) \widetilde\Psi(p) = 
\widetilde\Psi(p - q) \ .
\label{momentumshift}
\end{equation}
For momentum eigenstates $\langle x|k\rangle = \exp(i k x)$ (which are 
orthonormalized to $\langle k'|k\rangle = 2 \pi \delta(k - k')$) we have 
\begin{equation}
\langle x|k + q\rangle = \exp(i q x) \langle x|k\rangle = 
\widetilde U_q \langle x|k\rangle \ .
\end{equation}
We also find
\begin{eqnarray}
U_a \widetilde U_q \Psi(x)&=&U_a \exp(i q x) \Psi(x) =
\exp(i q (x + a)) \Psi(x + a), \nonumber \\
\widetilde U_q U_a \Psi(x)&=&\exp(i q x) \Psi(x + a) \ .
\end{eqnarray}
Hence, as a counterpart to the Heisenberg algebra $[\hat x,\hat p] = i$, one 
obtains the Weyl group relation
\begin{equation}
U_a \widetilde U_q = \exp(i q a) \widetilde U_q U_a \ .
\label{Weylxp}
\end{equation}
The action of the operators $U_a$ and $\widetilde U_q$ is illustrated in 
Fig.\ref{Fig1}.

\subsection{Standard momentum measurements}

Since the standard momentum operator $\hat p = - i \p_x$ is self-adjoint only 
over the entire real axis but not over the half-line, applying the standard 
concept of a momentum measurement requires to extend the Hilbert space from 
$L^2(\R_{\geq 0})$ to $L^2(\R)$ \cite{Coh77}. This can be achieved by 
assigning a finite constant potential $V(x<0) = V_0$ to the negative real axis 
and finally sending $V_0 \rightarrow \infty$ \cite{Gar04}. A momentum 
measurement then projects the wave function (which is exponentially suppressed 
on the negative real axis) on a plane wave $\langle x|k\rangle = \exp(i k x)$, 
which is unsuppressed for $x < 0$. Such a momentum measurement catapults the 
particle out of the energetically allowed region and transfers an infinite 
amount of energy to the particle in the limit $V_0 \rightarrow \infty$. Such a
measurement can be realized, for example, in an optical box trap, if the
ultra-cold atoms are released from the trap immediately before the momentum
measurement.

Let us first consider the bound state $|\psi_b\rangle$ for $\gamma < 0$. Its
overlap with the eigenstates $|k\rangle$ determines the probability density to 
obtain the value $k$ in a measurement of the standard momentum operator
\begin{equation}
\frac{1}{2 \pi} |\langle k|\psi_b\rangle|^2 = 
- \frac{1}{\pi} \frac{\gamma}{\gamma^2 + k^2} \ .
\end{equation}
As expected, the resulting momentum expectation value vanishes. The momentum 
uncertainty diverges because
\begin{equation}
\frac{1}{2 \pi} \int_{- \infty}^\infty dk \ k^2 |\langle k|\psi_b\rangle|^2 = 
- \frac{1}{\pi} \int_{- \infty}^\infty dk \ \frac{\gamma k^2}{\gamma^2 + k^2} =
\infty \ .
\end{equation}
In the limit $\gamma \rightarrow 0^-$ the bound state becomes an unbound 
zero-energy scattering state, and the probability density to measure the 
momentum $k$ turns into
\begin{equation}
- \lim_{\gamma \rightarrow 0^-} \frac{1}{\pi} \frac{\gamma}{\gamma^2 + k^2} =
\delta(k) \ .
\end{equation}
This seems to suggest that, in this limit, the value $k = 0$ is measured with 
certainty, in contradiction to the divergent uncertainty that we found for
$\gamma < 0$. This shows that the limit $\gamma \rightarrow 0^-$ is not
approached uniformly. In any case, the state $\psi_b(x) = A \theta(x)$, which 
in the limit $\gamma \rightarrow 0^-$ becomes proportional to the step function 
$\theta(x)$, is not identical with the $k = 0$ momentum eigenstate, which is 
constant over the entire real axis. This property results from the fact that 
$|\psi_b\rangle$ belongs to the Hilbert space $L^2(\R_{\geq 0})$ while 
$|k\rangle$ resides in an extension of $L^2(\R)$.

Next, we consider standard momentum measurements performed on the positive
energy scattering state $|\psi_E\rangle$ with $E = \tfrac{p^2}{2 m}$, 
$p \geq 0$. It is straightforward to obtain
\begin{equation}
\langle k|\psi_E\rangle = - i \lim_{\epsilon \rightarrow 0^+} \left(
\frac{1}{k - i \epsilon + p} + R(p) \frac{1}{k - i \epsilon - p} \right) \ .
\label{kpsiEoverlap}
\end{equation}
Using the residue theorem one then confirms that 
\begin{equation}
\langle \psi_{E'}|\psi_E\rangle = 
\frac{1}{2 \pi} \int_{- \infty}^\infty dk \ 
\langle \psi_{E'}|k\rangle \langle k|\psi_E\rangle = 2 \pi \delta(p - p') \ .
\end{equation}
In this case, one might expect that the only possible measurement results for 
the standard momentum are $k = p$ and $k = - p$, each with probability 
$\tfrac{1}{2}$. However, as we will see at the end of the paper, these 
probabilities are only $\tfrac{1}{4}$. Due to the two different Hilbert spaces 
$L^2(\R_{\geq 0})$ and $L^2(\R)$, $\psi_E(x)$, which vanishes for $x < 0$, is not 
a linear combination of $\langle x|- p\rangle = \exp(- i p x)$ and 
$R(p) \langle x|p\rangle = R(p) \exp(i p x)$ along the entire real axis.

We may conclude that it is possible to enforce the standard concept of momentum
for a particle on the half-line, however, at the price of putting the particle 
also onto the negative real axis as a result of the measurement. We will soon
present an alternative concept for a self-adjoint momentum operator, which 
actually leads to the same probability distribution for the measurement results,
but strictly confines the particle to the positive real axis also after a
measurement.

\subsection{Affine quantization on $\R_{\geq 0}$}

Since on the half-line there is no translation invariance, the operator 
$\hat p = - i \p_x$ is not self-adjoint. As a result, the standard commutation 
relation $[\hat x,\hat p] = i$ that underlies canonical quantization has only a 
formal status and is no longer physically meaningful. Consequently, in this 
case canonical quantization based upon $\hat p$ fails. For this reason, on the 
half-line affine quantization has played an important role 
\cite{Twa06,Kla12,Ber14,Alm18,Gou20}. The momentum operator $\hat p$ is then 
replaced by the generator $\hat d$ of infinitesimal dilations. The unitary 
operator that dilates a wave function by a scale factor $1/s \in \R_{>0}$ acts 
as
\begin{equation}
U_s \Psi(x) = \exp(i \hat d \log s) \Psi(x) = \sqrt{s} \Psi(s x) \ .
\label{scaleshift}
\end{equation}
Let us consider an infinitesimal dilation $s = 1 + \epsilon$ with small
$\epsilon$. We then obtain
\begin{eqnarray}
U_s \Psi(x)&=&\sqrt{1 + \epsilon} \Psi(x + \epsilon x) \nonumber \\
&\approx&\left(1 + \frac{\epsilon}{2}\right)
\left(\Psi(x) + \epsilon x \p_x \Psi(x)\right) \nonumber \\
&\approx&(1 + i \epsilon \hat d) \Psi(x) \ \Rightarrow \nonumber \\
\hat d \Psi(x)&=&- i \left(\frac{1}{2} + x \p_x\right) \Psi(x) \ \Rightarrow
\nonumber \\
\hat d&=&- i \left(\frac{1}{2} + x \p_x\right) = 
- \frac{i}{2} \left(\p_x x + x \p_x\right) \ .
\end{eqnarray}
The commutation relation that replaces $[\hat x,\hat p] = i$ in affine 
quantization is
\begin{equation}
[\hat x,\hat d] = [\hat x,- i x \p_x] = i \hat x \ .
\end{equation}

By performing a partial integration we obtain
\begin{eqnarray}
\langle \hat d^\dagger \chi|\Psi\rangle&=&\langle \chi|\hat d \Psi\rangle 
\nonumber \\
&=&\int_0^\infty dx \ \chi(x)^* \left[- i \left(\frac{1}{2} + x \p_x\right)
\Psi(x)\right] \nonumber \\
&=&\int_0^\infty dx \left[- i \left(\frac{1}{2} + x \p_x\right) \chi(x)\right]^*
\Psi(x) \nonumber \\
&=&\langle \hat d^\dagger \chi|\Psi\rangle \ .
\end{eqnarray}
As long as $x \chi(x)^* \Psi(x)$ vanishes at the boundary $x = 0$, $\hat d$ is 
Hermitean. Since square-integrable wave functions are less singular than 
$1/\sqrt{x}$ near the origin, this condition is satisfied without further
domain restrictions. As a result, $D(\hat d) = D(\hat d^\dagger)$ and $\hat d$ 
is indeed self-adjoint.

The eigenfunctions of $\hat d$ obey
\begin{eqnarray}
&&\hat d |\varkappa\rangle = 
- i \left(\frac{1}{2} + x \p_x\right) |\varkappa\rangle = 
\varkappa |\varkappa\rangle \ \Rightarrow \nonumber \\
&&\langle x|\varkappa\rangle = 
\frac{1}{\sqrt{l}} \left(\frac{x}{l}\right)^{i \varkappa - 1/2} \ .
\end{eqnarray}
Here $l$ is an arbitrarily chosen fixed length scale. Similar to the momentum 
eigenstates $|k\rangle$ over $\R$, the eigenstates $|\varkappa\rangle$ of 
$\hat d$ are not normalizable in the usual sense. The analog of Fourier 
transformation for canonical quantization is a Mellin-type transformation 
\cite{Twa06} for affine quantization
\begin{eqnarray}
\widetilde \Psi(\varkappa)&=&\langle \varkappa|\Psi\rangle = 
\int_0^\infty dx \ \langle \varkappa|x\rangle \langle x|\Psi\rangle \nonumber \\
&=&\int_0^\infty dx \ \frac{1}{\sqrt{l}} 
\left(\frac{x}{l}\right)^{- i \varkappa - 1/2} \Psi(x) \ .
\end{eqnarray}

Let us also introduce the unitary operator $\widetilde U_\delta =
\exp(i \delta \hat x)$ which leads to
\begin{eqnarray}
U_s \widetilde U_\delta \Psi(x)&=&U_s \exp(i \delta x) \Psi(x) =
\sqrt{s} \exp(i \delta s x) \Psi(s x) \ , \nonumber \\
\widetilde U_{\delta s} U_s \Psi(x)&=&\exp(i \delta s x) \sqrt{s} \Psi(s x) \ .
\end{eqnarray}
Hence, in analogy to the Weyl group for canonical quantization, for affine 
quantization one obtains the affine group relation
\begin{equation}
U_s \widetilde U_\delta = \widetilde U_{\delta s} U_s \ .
\end{equation}

\section{From a new concept for the momentum 
operator to canonical quantization on the half-line}

In this section, we introduce a new concept for the momentum operator in a
space with sharp boundaries, which allows us to apply canonical quantization
to the half-line.

\subsection{A self-adjoint momentum operator on the
half-line}

Recently we have developed a new concept for the momentum of a quantum
mechanical particle in a 1-d box $[-\tfrac{L}{2},\tfrac{L}{2}]$ \cite{AlH20},
which readily extends to the half-line as well as to higher-dimensional spaces
with sharp boundaries. For the convenience of the reader, we
summarize the most important aspects of this construction in Appendix A.
The construction results from the continuum limit of a system that is 
regularized on a spatial lattice. On a lattice, the derivative that enters the 
momentum operator is replaced by a nearest-neighbor finite difference. One must 
distinguish forward and backward derivatives, neither of them being Hermitean. 
Only the symmetrized forward-backward derivative, which corresponds to a 
next-to-nearest neighbor finite difference that extends over two lattice 
spacings, results in a Hermitean momentum operator. The lattice is naturally 
divided into two sublattices, one with even and one with odd lattice sites. The 
symmetrized forward-backward derivative associated with an even site then 
results from the values of the wave function at the two neighboring odd sites. 
In the continuum limit, the sublattice structure naturally leads to a 
two-component wave function, on which the momentum operator acts as a 
$2 \times 2$ matrix
\begin{equation}
\hat p_R = - i \left(\begin{array}{cc} 0 & \p_x \\ \p_x & 0 
\end{array}\right) = - i \sigma_1 \p_x, \
\Psi(x) = \left(\begin{array}{c} \Psi_e(x) \\ \Psi_o(x) \end{array}\right).
\label{2component}
\end{equation}
As a result, the problem is elevated to the Hilbert space $L^2(\R_{\geq 0}^2)$ of
square-integrable functions on the double-cover of the half-line. This is the
crucial insight that leads to the construction of a self-adjoint momentum 
operator. 

The full momentum operator $\hat p_R + i \hat p_I$ has both a Hermitean 
component $\hat p_R$ (which can be extended to a self-adjoint operator) and an 
anti-Hermitean component $i \hat p_I$ with
\begin{equation}
\hat p_I = \frac{1}{2} \lim_{\epsilon \rightarrow 0^+} \left(\begin{array}{cc}
\delta(x - \epsilon) & 0 \\ 0 & 0 \end{array}\right) \ .
\end{equation}
The operator $\hat p_I$ is self-adjoint and diagonal in the position basis. 

Let us first investigate the Hermiticity of $\hat p_R$. By partial integration 
we obtain
\begin{eqnarray}
&&\langle \hat p_R^\dagger \chi|\Psi\rangle = \langle \chi|\hat p_R \Psi\rangle =
\nonumber \\
&&\langle \hat p_R \chi|\Psi\rangle + i [\chi_e(0)^* \Psi_o(0) + 
\chi_o(0)^* \Psi_e(0)] \ .
\label{pRHermiticity}
\end{eqnarray}
Next, we impose the boundary condition
\begin{equation}
\Psi_o(0) = \lambda \Psi_e(0) \ ,
\label{momentumbc}
\end{equation}
which constrains the domain $D(\hat p_R)$. Inserting this relation in 
eq.(\ref{pRHermiticity}), the Hermiticity condition becomes
\begin{equation}
[\chi_e(0)^* \lambda + \chi_o(0)^*] \Psi_e(0) = 0 \ .
\end{equation}
Since $\Psi_e(0)$ can still take arbitrary values, one obtains
\begin{equation}
\chi_o(0) = - \lambda^* \chi_e(0) \ .
\end{equation} 
The operator $\hat p_R$ is self-adjoint if $D(\hat p_R^\dagger) = D(\hat p_R)$, 
which is true when $\lambda = - \lambda^*$, such that $\lambda \in i \R$. As a 
result, we obtain a 1-parameter family of self-adjoint extensions, characterized
by the purely imaginary parameter $\lambda$. Since $\hat p_R$ is self-adjoint, 
it has a complete set of orthonormal eigenstates with corresponding real 
eigenvalues. The momentum eigenstates, which obey 
$\hat p_R \phi_k(x) = k \phi_k(x)$ with $k \in \R$, are given by
\begin{eqnarray}
&&\phi_k(x) = \frac{1}{\sqrt{2}} \left(\begin{array}{c} 
\exp(i k x) + \sigma \exp(- i k x) \\ \exp(i k x) - \sigma \exp(- i k x) 
\end{array}\right) \ , \nonumber \\ 
&&\sigma = \frac{1 - \lambda}{1 + \lambda} \in U(1) \ .
\label{momentumeigenfunctions}
\end{eqnarray}
Like the momentum eigenstates $|k\rangle$ on the entire real axis, the states
$|\phi_k\rangle$ on the half-line are orthonormalized to $\delta$-functions, 
i.e.\ $\langle \phi_{k'}|\phi_k\rangle = 2 \pi \delta(k - k')$. 

\subsection{Embedding of $\hat H$ in $L^2(\R_{\geq 0}^2)$}

The search for a self-adjoint momentum operator that is strictly limited to the
half-line has naturally put us into the Hilbert space $L^2(\R_{\geq 0}^2)$ of the
doubly covered positive real axis. The double cover reflects the importance of 
even and odd lattice points in the underlying ultraviolet regularization at the 
level of the continuum description that emerges in the limit of vanishing 
lattice spacing. In order to apply the new concept for the momentum to the 
particle on the half-line, we must embed the original Hamiltonian 
$\hat H = - \tfrac{1}{2 m} \p_x^2 + V(x)$ with the self-adjoint extension 
parameter $\gamma$, which acts in $L^2(\R_{\geq 0})$, into the doubled Hilbert 
space $L^2(\R_{\geq 0}^2)$. This is achieved by constructing the Hamiltonian
\begin{equation}
\hat H(\mu) = \left(\begin{array}{cc} - \tfrac{1}{2 m} \p_x^2 + V(x) & 0 \\ 
0 & - \tfrac{1}{2 m} \p_x^2 + V(x) \end{array}\right) + \mu \hat P_- \ .
\label{doubledHamiltonian}
\end{equation}
Here $\hat P_-$ projects on states $\Psi^-(x)$ with 
$\Psi^-_o(x) = - \Psi^-_e(x)$. In the underlying lattice theory, these states 
have energies at the lattice cut-off. In order to decouple them from the 
continuum theory, we take the limit $\mu \rightarrow \infty$. The complementary 
operator $\hat P_+$ projects on the remaining states $\Psi^+(x)$ with 
$\Psi^+_o(x) = \Psi^+_e(x)$, which have finite energy, i.e.\
\begin{eqnarray}
&&\hat P_\pm = \frac{1}{2} 
\left(\begin{array}{cc} 1 & \pm 1 \\ \pm 1 & 1 \end{array}\right) \ , \quad
\hat P_\pm^2 = \hat P_\pm \ , \quad \hat P_+ + \hat P_- = \1 \ , \nonumber \\
&&\Psi(x) = \Psi^+(x) + \Psi^-(x) \ , \quad \Psi^\pm(x) = \hat P_\pm \Psi(x) \ .
\end{eqnarray}

What is the most general boundary condition at $x = 0$ for the extended 
Hamiltonian $\hat H(\mu)$? The linearity of quantum mechanics restricts us to 
write
\begin{equation}
\left(\begin{array}{c} \Psi_o(0) \\ 
\p_x \Psi_o(0) \end{array}\right) = e^{i \eta}
\left(\begin{array}{cc} a & - b \\ - c & d \end{array}\right) 
\left(\begin{array}{c} \Psi_e(0) \\ \p_x \Psi_e(0) \end{array}\right) \ .
\label{generalbc}
\end{equation}
Self-adjointness of $\hat H(\mu)$ again demands that the probability current
density vanishes at the origin, $j(0) = 0$. For 2-component wave functions the
current density takes the form
\begin{eqnarray}
j(x)&=&\frac{1}{2 m i} [\Psi(x)^* \p_x \Psi(x) - \p_x \Psi(x)^* \Psi(x)]
\nonumber \\
&=&\frac{1}{2 m i} [\Psi_e(x)^* \p_x \Psi_e(x) - \p_x \Psi_e(x)^* \Psi_e(x) 
\nonumber \\
&+&\Psi_o(x)^* \p_x \Psi_o(x) - \p_x \Psi_o(x)^* \Psi_o(x)] \ .
\end{eqnarray}
Using $j(0) = 0$, it is straightforward to derive the conditions 
$a, b, c, d \in \R$ and $a d - b c = - 1$. Together with $\eta$, these are
four independent parameters, which define a family of self-adjoint extensions.
In order to correctly embed the original Hamiltonian $\hat H$, the boundary 
condition must support the finite-energy states with 
$\Psi^+_o(x) = \Psi^+_e(x)$, which requires
\begin{equation}
e^{i \eta} = 1 \ , \quad a = 1 \ , \quad b = 0 \ , \quad d = - 1 \ .
\label{specialparameters}
\end{equation}
Using these specific parameters, eq.(\ref{generalbc}) reduces to
\begin{equation}
- \frac{c}{2} \Psi^+(0) - \p_x \Psi^+(0) = 0 \ , \quad \Psi^-(0) = 0 \ .
\label{specialbc}
\end{equation}
We now identify $\Psi^+(x)$ with the wave functions in the original Hilbert
space $L^2(\R_{\geq 0})$. Putting $\gamma = - c/2$, eq.(\ref{specialbc}) reduces
to the Robin boundary condition of eq.(\ref{Robinbc}), while the wave functions
$\Psi^-(x)$ obey Dirichlet boundary conditions. By construction, in the limit
$\mu \rightarrow \infty$ the Hamiltonian $\hat H(\mu)$ has the same 
finite-energy spectrum as the original Hamiltonian $\hat H$. The corresponding 
eigenstates of $\hat H(\mu)$ are just identical copies of the original 
eigenstates of $\hat H$ (renormalized by $\tfrac{1}{\sqrt{2}}$) in the upper 
and lower component of the 2-component wave function.

\subsection{Momentum measurements on $\R_{\geq 0}$}

We are now ready to apply the new concept of momentum to the particle on the
half-line. An original wave function $\psi(x) \in L^2(\R_{\geq 0})$ is trivially
embedded in the doubled Hilbert space $L^2(\R_{\geq 0})^2$ as
\begin{equation}
\Psi^+(x) = 
\left(\begin{array}{c} \Psi^+_e(x) \\ \Psi^+_o(x) \end{array}\right) =
\frac{1}{\sqrt{2}} \left(\begin{array}{c} \psi(x) \\ \psi(x) \end{array}\right)
\ .
\end{equation} 
The probability to measure a momentum value $k$ is determined by the amplitude
\begin{eqnarray}
\langle \phi_k|\Psi^+\rangle&=&\frac{1}{2} \! \int_0^\infty \!\!\!\! dx \!
\left(\begin{array}{c} 
\exp(i k x) + \sigma \exp(- i k x) \\ \exp(i k x) - \sigma \exp(- i k x) 
\end{array}\right)^{\!\dagger} \!\!
\left(\begin{array}{c} \psi(x) \\ \psi(x) \end{array}\right) \nonumber \\
&=&\int_{- \infty}^\infty dx \ \exp(- i k x) \psi(x) \theta(x) = 
\langle k|\psi\rangle \ .
\end{eqnarray}
Remarkably, $\langle k|\psi\rangle$ is just the amplitude that determines the
probability to obtain the value $k$ in a measurement of the standard momentum
operator $\hat p = - i \p_x$ that is self-adjoint only over the entire real 
axis. Here, using the step function $\theta(x)$, the wave function $\psi(x)$
has been trivially extended to the negative real axis.
As a result, both the standard and the new concept of momentum yield the 
same probability distributions for the measurement results. However, the two
concepts project onto different states after the measurements. In particular,
while a standard momentum measurement puts the particle also onto the negative
real axis, the new concept is strictly limited to the half-line, i.e.\ also 
after the measurement the particle remains on the positive real axis. Still,
also with the new concept a momentum measurement transfers an infinite amount
of energy to the particle. This is because the momentum eigenstate $\phi_k(x)$,
onto which a measurement that results in the value $k$ projects, not only has
the finite-energy component $\phi_k^+(x) = \tfrac{1}{\sqrt{2}} \exp(i k x)$ but 
also the component $\phi_k^-(x) = \tfrac{1}{\sqrt{2}} \sigma \exp(- i k x)$, 
whose energy diverges in the limit $\mu \rightarrow \infty$. Interestingly, the
purely imaginary self-adjoint extension parameter $\lambda$, which determines
$\sigma = (1 - \lambda)/(1 + \lambda) \in U(1)$, does not affect the 
probability to obtain a certain momentum measurement result.

As we will see later, the standard and the new concept for the 
momentum no longer result in the same measurement results when one considers
a particle that is confined to an interval \cite{AlH20}. In that case,
a standard momentum measurement catapults the particle outside of the interval 
and results in a continuous momentum value. The new concept, on the other hand,
yields discrete momentum values and leaves the particle inside the interval 
after a measurement.

\subsection{Canonical quantization on the half-line 
$\R_{\geq 0}$}

As we have seen in eq.(\ref{coordinateshift}), in canonical quantization the 
unitary operator $U_a = \exp(i \hat p a)$ with $\hat p = - i \p_x$ performs a 
coordinate shift, $U_a \Psi(x) = \Psi(x + a)$, while the operator 
$\widetilde U_q = \exp(i q \hat x)$ with $\hat x = i \p_p$ performs a momentum 
shift, $\widetilde U_q \widetilde\Psi(p) = \widetilde\Psi(p - q)$ (cf.\ 
eq.(\ref{momentumshift})). What is the unitary operator $\widetilde V_q$ that 
shifts the eigenstates $\phi_k(x)$ of the self-adjoint momentum operator on the 
half-line, $\hat p_R = - i \sigma_1 \p_x$, to $\phi_{k + q}(x)$? We construct
\begin{eqnarray}
&&\phi_{k + q}(x) = \frac{1}{\sqrt{2}} \left(\begin{array}{c} 
\exp(i (k + q) x) + \sigma \exp(- i (k + q) x) \\
\exp(i (k + q) x) - \sigma \exp(- i (k + q) x) \end{array}\right) \nonumber \\
&&=\! \left(\begin{array}{cc} \cos(q x) & i \sin(q x) \\ i \sin(q x) & \cos(q x)
\end{array}\right) \! \frac{1}{\sqrt{2}} \! \left(\begin{array}{c}
\exp(i k x) + \sigma \exp(- i k x) \\
\exp(i k x) - \sigma \exp(- i k x) \end{array}\right) \nonumber \\
&&= \widetilde V_q \phi_k(x) \ \Rightarrow \nonumber \\
&&\widetilde V_q = 
\left(\begin{array}{cc} \cos(q x) & i \sin(q x) \\ i \sin(q x) & \cos(q x)
\end{array}\right) = \exp(i q \hat x_R), \hat x_R = \sigma_1 x.
\label{momentumshifthalfline}
\end{eqnarray}
Not surprisingly, the conjugate coordinate to the momentum 
$\hat p_R = - i \sigma_1 \p_x$ is $\hat x_R = \sigma_1 x$. 
Indeed, the two operators obey the canonical commutation relation 
$[\hat x_R,\hat p_R] = i$. According to the Stone-von-Neumann theorem 
\cite{Sto30,Neu31,Neu32,Sto32}, this implies that $\hat x_R$ and $\hat p_R$ are 
unitarily equivalent to the standard coordinate $\hat x$ and momentum $\hat p$. 
While this is true mathematically, if not properly interpreted, it might 
be physically misleading. In particular, while the eigenstates of $\hat x$ 
describe positions on the entire real axis, the eigenstates of $\hat x_R$ 
describe a double-cover of the half-line. Hence, the situations are, in fact,
physically distinct, but, as we will see explicitly below, indeed mathematically
related by a unitary transformation. This may still seem surprising, since 
$\hat p_R$ possesses the self-adjoint extension parameter $\lambda \in i \R$ 
which characterizes the domain $D(\hat p_R)$, while the standard momentum 
operator does not. However, two operators $\hat p_R$, which are endowed with 
different self-adjoint extension parameters $\lambda$ and $\lambda'$, are 
indeed related by a unitary transformation
\begin{eqnarray}
&&W = \exp(i \omega \sigma_1) = 
\left(\begin{array}{cc} \cos\omega & i \sin\omega \\ i \sin\omega & \cos\omega
\end{array}\right) \ , \nonumber \\
&&W (- i \sigma_1 \p_x) W^\dagger = - i \sigma_1 \p_x \ ,
\label{Wequation}
\end{eqnarray}
which leaves the differential expression for $\hat p_R$ invariant. The boundary
condition $\Psi_o(0) = \lambda \Psi_e(0)$ then turns into
\begin{eqnarray}
\left(\begin{array}{c} \Psi'_e(0) \\ \Psi'_o(0) \end{array}\right)\!\!&=&\!
W\!\!\left(\begin{array}{c} \Psi_e(0) \\ \Psi_o(0) \end{array}\right)\!\!=\!\!
\left(\begin{array}{c} \cos\omega \Psi_e(0) + i \sin\omega \Psi_o(0) \\
i \sin\omega \Psi_e(0) + \cos\omega \Psi_o(0) \end{array}\right) \nonumber \\
&=&\left(\begin{array}{c} (\cos\omega + \lambda i \sin\omega) \Psi_e(0) \\
(i \sin\omega + \lambda \cos\omega) \Psi_e(0) \end{array}\right) \
\Rightarrow \nonumber \\
\Psi'_o(0)&=& 
\frac{i \sin\omega + \lambda \cos\omega}{\cos\omega + \lambda i \sin\omega}
\Psi'_e(0) = \lambda' \Psi'_e(0) \ .
\label{uequivalence}
\end{eqnarray}
As a result, two different operators $\hat p_R$, which are associated with the 
self-adjoint extension parameters $\lambda$ and $\lambda'$ are related by the
unitary transformation $W$ with
\begin{eqnarray}
&&\tan\omega = i \frac{\lambda - \lambda'}{1 - \lambda \lambda'} \ ,
\nonumber \\
&&\sigma' = \frac{1 - \lambda'}{1 + \lambda'} = 
\exp(- 2 i \omega) \frac{1 - \lambda}{1 + \lambda} = \exp(- 2 i \omega) \sigma
\ . 
\label{sigmatransformation}
\end{eqnarray}

Let us again denote the eigenstates of $\hat x$ as $|x\rangle$, such that 
$\hat x |x\rangle = x |x\rangle$. Here we limit ourselves to the positive real
axis with $x > 0$. The eigenstates of $\hat x_R = \sigma_1 x$ are then given by
\begin{eqnarray}
&&\hat x_R |x,+\rangle = x |x,+\rangle \ , \
|x,+\rangle = \frac{1}{\sqrt{2}} \left(\begin{array}{c} |x\rangle \\ |x\rangle
\end{array}\right) \ , \nonumber \\
&&\hat x_R |x,-\rangle = - x |x,-\rangle \ , \
|x,-\rangle = \frac{1}{\sqrt{2}} \left(\begin{array}{c} |x\rangle \\ - |x\rangle
\end{array}\right) \ .
\label{xReigenstates}
\end{eqnarray}
The eigenstates $|x,+\rangle$ have a positive eigenvalue $x > 0$ and belong
to the finite-energy sector, i.e.\ $\hat P_+ |x,+\rangle = |x,+\rangle$, while
the eigenstates $|x,-\rangle$ have a negative eigenvalue $- x$ and obey
$\hat P_- |x,-\rangle = |x,-\rangle$, $\hat P_+ |x,-\rangle = 0$.

\begin{figure}[tbp]
\vspace{3cm}
\begin{tikzpicture}
\tikz\draw
[line width=0.4mm,color=black] (4,0) -- (8,0)
node[right] {$\hat x_R$}
[line width=0.4mm,color=black] (7.9,0.1) -- (8,0)
[line width=0.4mm,color=black] (7.9,-0.1) -- (8,0)
[line width=0.4mm,color=black] (4,0) -- (4,-0.3)
node[left] {$0$}
[line width=0.4mm,color=black] (4,-0.3) -- (8,-0.3)
[line width=0.4mm,color=black] (7.9,-0.2) -- (8,-0.3)
[line width=0.4mm,color=black] (7.9,-0.4) -- (8,-0.3)
[line width=0.4mm,color=black] (6,-0.2) -- (6,0.2)
node[above] {$|x,+\rangle$}
[line width=0.4mm,color=black] (5,-0.5) -- (5,-0.1)
[line width=0.4mm,color=black] (6,0.3) -- (4,0.3)
node[above] {$V_a$}
[line width=0.4mm,color=black] (4,-0.6) -- (5,-0.6)
node[below] {$|a - x,-\rangle$}
[line width=0.4mm,color=black] (4,0.3) arc (90:270:0.45)
[line width=0.4mm,color=black] (4.9,-0.5) -- (5,-0.6)
[line width=0.4mm,color=black] (4.9,-0.7) -- (5,-0.6)
[line width=0.4mm,color=black] (0,-2) -- (8,-2)
node[right] {$\hat p_R$}
[line width=0.4mm,color=black] (7.9,-1.9) -- (8,-2)
[line width=0.4mm,color=black] (7.9,-2.1) -- (8,-2)
[line width=0.4mm,color=black] (5,-2.2) -- (5,-1.8)
node[above] {$|k + q\rangle$}
[line width=0.4mm,color=black] (2.5,-2.2) -- (2.5,-1.8)
node[above] {$|k\rangle$}
[line width=0.4mm,color=black] (2.5,-2.5) -- (5,-2.5)
node[below] {\hspace{-2.5cm} $\widetilde V_q$}
[line width=0.4mm,color=black] (4.9,-2.4) -- (5,-2.5)
[line width=0.4mm,color=black] (4.9,-2.6) -- (5,-2.5);
\end{tikzpicture}
\vspace{-3.5cm}
\caption{\it Action of the translation operators $V_a = \exp(i \hat p_R a)$ and
$\widetilde V_q = \exp(i q \hat x_R)$ on the position and momentum eigenstates 
for the half-line $\R_{\geq 0}$.}
\label{Fig2}
\end{figure}
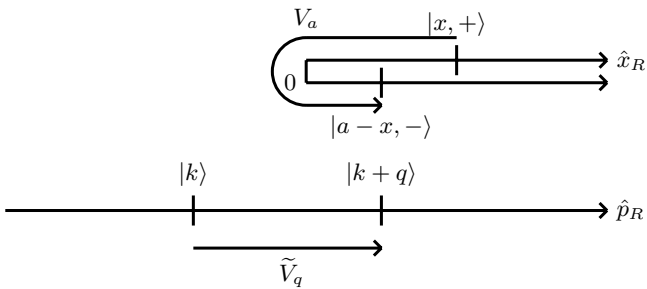

How does the operator $V_a = \exp(i \hat p_R a)$ act on the eigenstates
$|x,\pm\rangle$? We obtain
\begin{eqnarray}
V_a |x,\pm\rangle&=&\frac{1}{2 \pi} \int_{- \infty}^\infty dk 
\exp(i \hat p_R a) |\phi_k\rangle \langle \phi_k|x,\pm\rangle \nonumber \\
&=&\frac{1}{2 \pi} \int_{- \infty}^\infty dk \exp(i k a) |\phi_k\rangle 
\langle \phi_k|x,\pm\rangle \ , \nonumber \\
\langle x,+|\phi_k\rangle&=&\frac{1}{2} (1,1) \left(\begin{array}{c}
\exp(i k x) + \sigma \exp(- i k x) \\
\exp(i k x) - \sigma \exp(- i k x) \end{array}\right) \nonumber \\
&=&\exp(i k x) \ ,
\nonumber \\
\langle x,-|\phi_k\rangle&=&\frac{1}{2} (1,- 1) \left(\begin{array}{c}
\exp(i k x) + \sigma \exp(- i k x) \\
\exp(i k x) - \sigma \exp(- i k x) \end{array}\right) \nonumber \\
&=&\sigma \exp(- i k x) \ ,
\end{eqnarray}
which implies
\begin{eqnarray}
&&V_a |x,+\rangle = |x - a,+\rangle \ , \ 
\mbox{for} \ 0 \leq x - a \ , \nonumber \\
&&V_a |x,+\rangle = \sigma |a - x,-\rangle \ , \ 
\mbox{for} \ x - a < 0 \ , \nonumber \\
&&V_a |x,-\rangle = |x + a,-\rangle \ , \ 
\mbox{for} \ 0 \leq x + a \ , \nonumber \\
&&V_a |x,-\rangle = \sigma^* |- x - a,+\rangle \ , \ 
\mbox{for} \ x + a < 0 \ .
\label{positionshift}
\end{eqnarray}
Writing $\hat x_R |x_R\rangle = x_R |x_R\rangle$, $x_R \in \R$, we identify 
$|x_R\rangle = |x_R,+\rangle$ for $x_R > 0$, and $|x_R\rangle = |- x_R,-\rangle$
for $x_R < 0$. Eq.(\ref{positionshift}) then implies that, up to a phase 
$\sigma$ or $\sigma^*$, $V_a$ shifts the eigenstates of $\hat x_R$ from 
$|x_R\rangle$ to $|x_R - a\rangle$. Independent of $\sigma$, as a consequence 
of the Heisenberg algebra $[\hat x_R,\hat p_R] = i$, one obtains the Weyl group 
relation
\begin{equation}
V_a \widetilde V_q = \exp(i q a) \widetilde V_q V_a \ .
\label{WeylxRpR}
\end{equation}
The action of the operators $V_a$ and $\widetilde V_q$ is illustrated in 
Fig.\ref{Fig2}. The operator $V_a = \exp(i \hat p_R a)$ transports states 
$|x,+\rangle$ in the finite-energy sector to the left, and states $|x,-\rangle$,
whose energy is ultraviolet-sensitive, to the right. When a state $|x,+\rangle$ 
is transported beyond the origin (at $x = 0$), it turns into the state 
$|a - x,-\rangle$.

For mathematical purposes, the Weyl groups are more convenient than the
corresponding Heisenberg algebras because unitary operators are bounded while
the corresponding self-adjoint operators are not. Based upon the Weyl groups 
of eq.(\ref{Weylxp}) and eq.(\ref{WeylxRpR}), the Stone-von-Neumann theorem 
guarantees the unitary equivalence of $\hat x_R$ and $\hat p_R$ with $\hat x$ 
and $\hat p$. The corresponding unitary transformation $U$ is given by
\begin{eqnarray}
&&U V_a U^\dagger = U_a \ , \quad U \widetilde V_q U^\dagger = \widetilde U_q \ ,
\nonumber \\ 
&&U \hat x_R U^\dagger = \hat x \ , \quad U |x,+\rangle = |x\rangle \ , 
\quad U |x,-\rangle = \sigma^* |- x\rangle \ , 
\nonumber \\
&&U \hat p_R U^\dagger = \hat p \ , \quad U|\phi_k\rangle = |k\rangle \ .
\label{Uhalflineaxis}
\end{eqnarray}
We like to stress again that this mathematical unitary transformation relates
two physically very different situations. In particular, while the states 
$|x,+\rangle$ (with $x > 0$) in the finite-energy sector are mapped to position 
eigenstates $|x\rangle$ on the positive real axis, the states $|x,-\rangle$ 
(again with $x > 0$), which are mapped to the states $|- x\rangle$ on the 
negative real axis, actually reside on the positive half-line, but have 
ultraviolet-sensitive energies and are removed from the spectrum of the 
Hamiltonian $\hat H(\mu)$ in the limit $\mu \rightarrow \infty$.

\subsection{The classical limit}

Until now we have used the term quantization --- be it canonical or affine --- 
without even starting from a classical theory. Since quantum physics is more 
fundamental than classical physics, this way to proceed is not unreasonable. 
Until now, we have shown that our new concept gives rise to a self-adjoint 
momentum operator $\hat p_R = - i \sigma_1 \p_x$ that obeys the canonical 
commutation relation, $[\hat x_R,\hat p_R] = i$, with the corresponding 
coordinate $\hat x_R = \sigma_1 x$. We now consider the classical limit in 
order to demonstrate that the new concept still makes sense when commutators 
are replaced by Poisson brackets.

Let us perform a unitary transformation $D$ that diagonalizes $\hat x_R$, 
$\hat p_R$, and $\hat H(\mu)$ such that
\begin{eqnarray}
&&D \sigma_1 D^\dagger = \sigma_3 \ , \quad D = \frac{1}{\sqrt{2}}
\left(\begin{array}{cc} 1 & 1 \\ 1 & - 1 \end{array}\right) \ , \nonumber \\
&&\hat x_R' = D \hat x_R D^\dagger = x \sigma_3 = 
\left(\begin{array}{cc} x & 0 \\ 0 & - x \end{array}\right) \ , \quad 
x \in \R_{\geq 0} \ , \nonumber \\ 
&&\hat p_R' = D \hat p_R D^\dagger = - i \sigma_3 \p_x = - i
\left(\begin{array}{cc} \p_x & 0 \\ 0 & - \p_x \end{array}\right) \ , 
\nonumber \\ 
&&\hat H(\mu)' = D \hat H(\mu) D^\dagger \nonumber \\
&&= \left(\begin{array}{cc} - \tfrac{1}{2 m} \p_x^2 + V(x) & 0 \\ 
0 & - \tfrac{1}{2 m} \p_x^2 + V(x) + \mu \end{array}\right) \ .
\end{eqnarray}
The positive eigenvalues of $\hat x_R$ correspond to physical points 
$x \in \R_{\geq 0}$ on the half-line on which finite-energy states are 
located. The negative eigenvalues $- x$ of $\hat x_R$, on the other hand, are
not associated with the negative real axis, but host states with energies at
the ultraviolet cut-off scale, which are removed in the limit 
$\mu \rightarrow \infty$.

A similar mathematical description arises when one applies the standard 
concept of momentum. In that case, one regularizes the problem by assigning a
potential  $V(x<0) = V_0$ to points on the negative real axis, and ultimately
taking the limit $V_0 \rightarrow \infty$. At a formal level $V_0$ then
corresponds to $\mu$, but the physical interpretation is very different. In
fact, when one applies the new concept for the momentum, the particle is 
strictly limited to the positive real axis, even after a momentum measurement.

In any case, one can now replace the commutation relation 
$[\hat x_R',\hat p_R'] = i$ by the classical Poisson bracket relation
$\{x,p\} = 1$ and replace the Hamiltonian $\hat H(\mu)'$ in the limit 
$\mu \rightarrow \infty$ by the classical Hamilton function 
${\cal H}(x,p) = \tfrac{p^2}{2 m} + V(x)$. The hard wall at $x = 0$ arises as a
consequence of $\mu \rightarrow \infty$, thus leading to the standard motion of
a classical particle that gets reflected at the origin. The self-adjoint 
extension parameters $\lambda$ and $\gamma$ leave no trace in the classical 
limit. In particular, unlike the quantum particle, the classical particle 
cannot be bound to the reflecting wall at $x = 0$.
                 
\subsection{Subtleties with commutators and operator 
domains}

Canonical quantization (as defined in the introduction) usually starts from a 
classical theory with the Poisson bracket relation $\{x,p\} = 1$, which is then 
promoted to the commutation relation $[\hat x,\hat p] = i$ between self-adjoint 
operators $\hat x$ and $\hat p$. Until now, it was assumed that on the 
half-line canonical quantization fails, simply because in this case no 
self-adjoint momentum operator exists for $L^2(\R_{\geq 0})$. As we have seen, a 
satisfactory self-adjoint momentum operator $\hat p_R = - i \sigma_1 \p_x$ as 
well as its canonically conjugate coordinate $\hat x_R = \sigma_1 x$ (with 
$x \geq 0$) indeed exist in the doubled Hilbert space $L^2(\R_{\geq 0}^2)$. The 
operator $\hat p_R$ is endowed with the self-adjoint extension parameter 
$\lambda$, and is thus not uniquely defined. However, as we have seen in 
eq.(\ref{uequivalence}), in accordance with the Stone-von-Neumann theorem 
\cite{Sto30,Neu31,Neu32,Sto32}, two operators $\hat p_R$, which are associated 
with two different self-adjoint extension parameters $\lambda$ and $\lambda'$, 
are, in fact, unitarily equivalent.

Following the canonical quantization procedure, one replaces the classical 
Hamilton function ${\cal H}(x,p) = \tfrac{p^2}{2 m} + V(x)$ with a Hamilton 
operator $\hat H$ by replacing $x$ with $\hat x$ and $p$ with $\hat p$. When 
one replaces the classical kinetic energy ${\cal T} = \tfrac{p^2}{2 m}$ by
the operator \begin{equation}
\hat T = \frac{\hat p_R^2}{2 m} = \frac{1}{2 m} (- i \sigma_1 \p_x)^2 =
- \frac{1}{2 m} \p_x^2 \1 \ ,
\end{equation}
at least at the formal level of differential expressions one obtains the
correct kinetic energy operator that enters the Hamiltonian $\hat H(\mu)$
of eq.(\ref{doubledHamiltonian}). 

At a superficial level one might conclude that $\hat T$ and $\hat p_R$ commute.
However, $\hat T$ is self-adjoint in $L^2(\R_{\geq 0}^2)$ only if it is 
endowed with its own self-adjoint extension parameters which characterize the
domain $D(\hat T)$. The most general self-adjoint extension of $\hat T$ is
characterized by eq.(\ref{generalbc}), which contains four independent
parameters: $\eta, a, b, c, d \in \R$ subject to the constraint
$ad - bc = - 1$. However, for the reasons explained above, here we are only
interested in the special case $\eta = 0$, $a = 1$, $b = 0$, $c = - 2 \gamma$,
$d = - 1$. For general finite $\gamma < \infty$, one then has 
$\Psi_e(0) = \Psi_o(0)$, which is inconsistent with 
$\Psi_o(0) = \lambda \Psi_e(0)$, because $\lambda \in i \R$ is purely imaginary.
As a result, the domains of the momentum operator and the kinetic energy 
operator are not the same, $D(\hat p_R) \neq D(\hat T)$. Only when we choose 
Dirichlet boundary conditions (which correspond to $\gamma \rightarrow \infty$),
the domain of $\hat T$ is characterized by $\Psi_e(0) = \Psi_o(0) = 0$, which 
is automatically consistent with $\Psi_o(0) = \lambda \Psi_e(0)$. Hence, in 
this special case, $D(\hat T) \subset D(\hat p_R)$. This implies that an 
application of $\hat p_R$ on a wave function $\hat T \Psi(x) \in D(\hat T)$ is
still possible and $\hat p_R \hat T \Psi(x) \in D(\hat p_R)$. On the other hand,
an application of $\hat T$ on $\hat p_R \Psi(x) \in D(\hat p_R)$ is possible
only if $\hat p_R \Psi(x) \in D(\hat T)$. Hence, the commutator 
$\hat p_R \hat T - \hat T \hat p_R$ cannot act on all wave functions in
$D(\hat p_R)$ or $D(\hat T)$, and hence has only a limited formal meaning. As
a result, although the operators $\hat p_R$ and $\hat T$ seem to commute at the
superficial level of differential expressions, they do not have common
eigenfunctions.

Let us return to the operators $\hat x$ and 
$\hat d = - i (\tfrac{1}{2} + x \p_x)$ that we encountered in affine 
quantization and consider them together with the operators 
$\hat x_R = \sigma_1 x$ and $\hat p_R = - i \sigma_1 \p_x$ that we used in 
canonical quantization. Among those, only the domain $D(\hat p_R)$ is further 
restricted by the condition $\Psi_o(0) = \lambda \Psi_e(0)$ with 
$\lambda \in i \R$. First of all,
\begin{equation}
\hat x_R \Psi(x) = \left(\begin{array}{cc} 0 & x \\ x & 0 \end{array}\right)
\left(\begin{array}{c} \Psi_e(x) \\ \Psi_o(x) \end{array}\right) =
\left(\begin{array}{c} x \Psi_o(x) \\ x \Psi_e(x) \end{array}\right) \ ,
\end{equation}
which does not lead out of $D(\hat p_R)$, because 
$x \Psi_e(x) = \lambda x \Psi_o(x) = 0$ at $x = 0$. Trivially embedding
$\hat x$ into the doubled Hilbert space, we also obtain
\begin{equation}
\left(\begin{array}{cc} \hat x & 0 \\ 0 & \hat x \end{array}\right)
\left(\begin{array}{c} \Psi_e(x) \\ \Psi_o(x) \end{array}\right) =
\left(\begin{array}{c} x \Psi_e(x) \\ x \Psi_o(x) \end{array}\right) \ ,
\end{equation}
which does not lead out of $D(\hat p_R)$ either. Finally, by also embedding
$\hat d$ into the doubled Hilbert space, one obtains
\begin{equation}
\left(\begin{array}{cc} \hat d & 0 \\ 0 & \hat d \end{array}\right)
\left(\begin{array}{c} \Psi_e(x) \\ \Psi_o(x) \end{array}\right) =
\left(\begin{array}{c} - i (\tfrac{1}{2} + x \p_x) \Psi_e(x) \\ 
- i (\tfrac{1}{2} + x \p_x) \Psi_o(x) \end{array}\right) \ .
\end{equation}
Putting $x = 0$, we then find
\begin{eqnarray}
- i (\tfrac{1}{2} + x \p_x) \Psi_o(0)&=&- \tfrac{i}{2} \Psi_o(0) =
- \tfrac{i}{2} \lambda \Psi_e(0) \nonumber \\
&=&- \lambda i (\tfrac{1}{2} + x \p_x) \Psi_e(0) \ .
\end{eqnarray}
We thus conclude that $\hat d$ does not lead out of $D(\hat p_R)$ either. Since
$\hat x$, $\hat x_R$, and $\hat d$ are not subject to further domain 
restrictions of their own, limiting ourselves to wave functions 
$\Psi(x) \in D(\hat p_R)$, the following commutators are indeed not just
formal expressions, but mathematically and physically meaningful
\begin{eqnarray}
&&[\hat x,\hat x_R] = 0 \ , \quad [\hat x,\hat d] = i \hat x \ , \quad
[\hat x,\hat p_R] = i \sigma_1 \ , \nonumber \\
&&[\hat x_R,\hat d] = [\sigma_1 x,- i x \p_x] = i \sigma_1 x = i \hat x_R \ , 
\quad [\hat x_R,\hat p_R] = i \ , \nonumber \\
&&[\hat p_R,\hat d] = [- i \sigma_1 \p_x,- i x \p_x] = - \sigma_1 \p_x = 
- i \hat p_R \ .
\end{eqnarray}
This implies that affine and canonical quantization are fully consistent with 
each other and can be implemented simultaneously.

\subsection{Summary of the canonical quantization 
procedure}

To summarize, the resulting canonical quantization procedure consists of the 
following steps: 
\begin{itemize}
\item
Select a classical system characterized by coordinates $x$ and canonically
conjugate momenta $p$ associated with a classical Hamilton function 
${\cal H}(x,p)$ as well as with some observables ${\cal A}(x,p)$.
\item
Identify an appropriate Hilbert space in which to realize the time-evolution
and the measurements for a corresponding quantum system. 
\item
Replace classical Poisson bracket relations by formal commutation relations.
\item
Realize the commutation relations by differential expressions.
\item
Extend the differential expressions to self-adjoint operators.
\item
Make a specific choice of self-adjoint extension parameters and thus of the
corresponding operator domains.
\item
Return to the formal commutation relations, investigate whether they are
affected by domain incompatibilities, and take this into account properly.
\item
If the canonical commutation relation between position and momentum itself is
compromised by domain incompatibilities, try to replace the position or the
momentum operator by more appropriate operators that are unaffected by such 
subtleties.    
\end{itemize}
As we have seen, the identification of an appropriate Hilbert space may be
non-trivial and may depend on the set of observables to be measured. 
In particular, for the particle on the half-line the time-evolution driven by 
the Hamiltonian $\hat H$ can be realized in the Hilbert space $L^2(\R_{\geq 0})$.
However, in order to consider momentum measurements, the Hilbert space must be 
doubled to $L^2(\R_{\geq 0}^2)$ and the Hamiltonian must be extended accordingly 
to $\hat H(\mu)$. In the absence of subtle Hilbert space or domain 
incompatibility issues, all relevant commutators correctly reflect the 
relations between the various operators, and the procedure outlined above
simplifies considerably. When such subtleties do arise, on the other hand, 
naively applying the commutation relations that result from classical Poisson
brackets may lead to wrong results. This would happen, for example, if one 
would not realize that the self-adjoint operators $\hat p_R$ and $\hat T$ do 
not commute, although the corresponding differential expressions seem to 
suggest this. In such a case, it is necessary to properly address the 
subtleties, which is possible using the well-developed mathematical theory that 
was initiated by von Neumann \cite{Neu32a}.

The remaining steps for addressing the quantum dynamics are the usual ones, but 
they are strongly affected by the selected operator domains:
\begin{itemize}
\item
Employ the Hamiltonian to determine the wave function that solves the 
Schr\"odinger equation.
\item
Solve the eigenvalue problem for the operators that describe observables to be
measured.
\item
Project the wave function on the eigenfunction of an observable in order to
predict the probability to measure the corresponding eigenvalue.
\end{itemize}
It should be stressed that there is absolutely nothing wrong with affine 
quantization applied to the half-line. In view of our construction, it is just
no longer the only quantization procedure that is available in this case.

\subsection{Is canonical quantization a fine quantization for 
the half-line?}

Is there anything wrong with the above canonical quantization procedure applied
to the half-line? We will now argue that this admittedly somewhat subtle 
procedure is completely fine, and, in fact, very natural in a continuous space 
with sharp boundaries. First of all, in order to motivate the construction of 
$\hat p_R$ in the first place, in \cite{AlH20} we have introduced an ultraviolet
regularization by replacing the continuum with a spatial lattice with $N$ 
points. The resulting Hilbert space is then $N$-dimensional. In a 
finite-dimensional Hilbert space, there is no difference between Hermiticity 
and self-adjointness, and one need not worry about the domains of operators. 
The lattice variant of the momentum operator $\hat p_R$ results from a 
symmetrized forward-backward lattice derivative that extends over two lattice 
spacings. The lattice variant of the kinetic energy operator $\hat T$ is 
described by a standard finite-difference lattice version of a second 
derivative. The low-energy physics of the lattice system does not require an 
extension of the Hilbert space and does not depend on any operator domain 
issues. If we interpret the lattice spacing as mimicking a shortest relevant 
physical length scale, such as a crystal lattice spacing in a quantum dot, or 
the Planck length in an interval of extra-dimensional space, every Hermitean 
operator would automatically be self-adjoint and could act in the entire 
Hilbert space. In particular, the mathematical subtleties related to operator 
domains would have no effect on the physics. 

However, also from a physics point of view, it is desirable to derive a 
low-energy effective description of the underlying lattice physics, by taking
the continuum limit of vanishing lattice spacing. This would have been 
straightforward if we would care only about the spectrum of the Hamiltonian.
In fact, one could then stay within the Hilbert space $L^2(\R_{\geq 0})$.
Since we also care about the momentum operator, the situation is more subtle.
In particular, the crucial concept of even and odd lattice points naturally 
leads to the doubled Hilbert space $L^2(\R_{\geq 0})$. As we have seen, the 
embedding of the original Hamiltonian $\hat H$ into this framework leads to
the Hamiltonian $\hat H(\mu)$ with the specific self-adjoint extension 
parameters of eq.(\ref{specialparameters}). The limit $\mu \rightarrow \infty$
removes those states from the finite-energy spectrum whose lattice variants have
energies at the cut-off scale. This inevitably leads to the different operator
domains, $D(\hat p_R) \neq D(\hat H(\mu))$. 

While the resulting canonical quantization procedure may seem unnecessarily 
complicated, it is the price a physicist has to pay for using a low-energy 
continuum description in an infinite-dimensional Hilbert space. This 
description is, in fact, mathematically quite elegant, but admittedly not 
completely straightforward. A straightforward alternative description could 
stay on the lattice without taking the continuum limit. However, such a 
formulation is not very transparent, it leads to more complicated variants of 
the Schr\"odinger equation, and it does not correspond to the standard quantum 
mechanical continuum description. We suggest that it is worth familiarizing 
oneself with the differences between Hermiticity and self-adjointness, and thus 
with the issues related to operator domains, even if this may require an 
expansion of the sometimes prevailing practices in quantum mechanics. 

\section{Canonical quantization in an interval and on a 
circle}

In this section, we apply the new concept for the momentum to an interval,
which allows us to extend canonical quantization to that case as well. We then
relate the results to the well-understood situation on a circle.

\subsection{Canonical quantization in the interval $[0,L]$}

In \cite{AlH20} we have introduced the new concept for the momentum operator
in an interval $[-\tfrac{L}{2},\tfrac{L}{2}]$. In that work we have only
briefly touched upon the corresponding algebra that relates the finite-volume
momentum to its corresponding conjugate coordinates. Here we consider this
problem in more detail. We shift the interval to $[0,L]$ in order to be able to
recover the half-line in the limit $L \rightarrow \infty$.

In an interval, the domain $D(\hat p_R)$ of the finite-volume momentum
operator $\hat p_R$ is characterized by two self-adjoint extension parameters,
$\lambda$ and $\lambda_L$, associated with the two ends of the interval. In the 
interval $[0,L]$ the boundary conditions are
\begin{equation}
\Psi_o(0) = \lambda \Psi_e(0) \ , \quad \Psi_o(L) = \lambda_L \Psi_e(L) \ ,
\quad \lambda, \lambda_L \in i \R_{\geq 0} \ .
\end{equation}
The corresponding momentum eigenfunctions $\phi_k(x)$ are still given by
eq.(\ref{momentumeigenfunctions}). However, in contrast to the half-line, in 
the interval the momentum $k$ is quantized and obeys the condition
\begin{equation}
\exp(2 i k L) = 
\frac{(1 - \lambda)(1 + \lambda_L)}{(1 + \lambda)(1 - \lambda_L)} = 
\sigma \sigma_L^* = \exp(i \theta) \ .
\label{thetaequation}
\end{equation}
For $\lambda = \lambda_L$ this leads to $k = \tfrac{\pi n}{L}$, $n \in \Z$, 
while in general
\begin{equation}
k = \frac{\pi}{L}\left(n + \frac{\theta}{2 \pi}\right) \ , \ n \in \Z \ .
\end{equation}
Adjacent momentum eigenvalues are hence separated by $\tfrac{\pi}{L}$. The
self-adjoint extension parameter $\theta$ is invariant against the unitary
transformation $W$ of eq.(\ref{Wequation}) because
\begin{eqnarray}
\exp(i \theta')&=&\sigma' {\sigma_L'}^* = 
\exp(- 2 i \omega) \sigma \exp(2 i \omega) \sigma_L^* \nonumber \\
&=&\sigma {\sigma_L}^* =
\exp(i \theta) \ .
\label{invarianttheta}
\end{eqnarray}
Here we have used eqs.(\ref{thetaequation}) and (\ref{sigmatransformation}).

What is the unitary operator that translates the momentum eigenfunctions by 
$\tfrac{\pi}{L}$? Using eq.(\ref{momentumshifthalfline}), which is still 
applicable in the interval, we obtain 
\begin{eqnarray}
&&\phi_{k+\pi/L}(x) = \widetilde V_{\pi/L} \phi_k(x) \ , \nonumber \\
&&\widetilde V_{\pi/L} = \exp\left(i \frac{\pi}{L} \hat x_R\right) =
\left(\begin{array}{cc} \cos\tfrac{\pi x}{L} & i \sin\tfrac{\pi x}{L} \\ 
i \sin\tfrac{\pi x}{L} & \cos\tfrac{\pi x}{L} \end{array}\right) \ .
\end{eqnarray}
The fact that $\widetilde V_{\pi/L}$ acts as a shift operator for the momentum
also reflects itself in the commutation relations
\begin{equation}
[\hat p_R,\widetilde V_{\pi/L}] = \frac{\pi}{L} \widetilde V_{\pi/L} \ , \quad
[\hat p_R,\widetilde V_{\pi/L}^\dagger] = 
- \frac{\pi}{L} \widetilde V_{\pi/L}^\dagger \ .
\label{pVcommutators}
\end{equation} 

Since for $x = 0$ and $x = L$ the operator $\widetilde V_{\pi/L}$ reduces to 
$\1$, for a wave function $\Psi(x) \in D(\hat p_R)$ (which obeys
$\Psi_o(0) = \lambda \Psi_e(0)$ and $\Psi_o(L) = \lambda_L \Psi_e(L)$) one 
obtains 
$(\widetilde V_{\pi/L} \Psi)_o(0) = \lambda (\widetilde V_{\pi/L} \Psi)_e(0)$ and
$(\widetilde V_{\pi/L} \Psi)_o(L) = \lambda_L (\widetilde V_{\pi/L} \Psi)_e(L)$. 
As a result, $\widetilde V_{\pi/L} \Psi(x) \in D(\hat p_R)$, which means that
$\widetilde V_{\pi/L}$ does not lead out of the domain $D(\hat p_R)$. This
implies that the commutation relations of eq.(\ref{pVcommutators}) are not just 
formal relations, but are fully consistent with the domains of the 
corresponding operators.

It should be noted that the canonical commutation relation 
$[\hat x_R,\hat p_R] = i$, which is completely appropriate on the half-line, is 
no longer a meaningful expression in the interval. This is because, in contrast
to $\widetilde V_{\pi/L}$, the operator $\hat x_R$ leads out of the domain
$D(\hat p_R)$. This follows from 
$(\hat x_R \Psi)_e(L) = L \Psi_o(L) = L \lambda_L \Psi_e(L) =
\lambda_L (\hat x_R \Psi)_o(L)$, which is inconsistent with the domain condition
$(\hat x_R \Psi)_o(L) = \lambda_L (\hat x_R \Psi)_o(L)$. Hence, in the interval 
$[0,L]$ the commutation relations of eq.(\ref{pVcommutators}) replace the 
canonical commutation relation $[\hat x_R,\hat p_R] = i$. This is another 
example where the last step in the canonical quantization procedure is crucial: 
``If the canonical commutation relation between position and momentum itself is 
compromised by domain incompatibilities, try to replace the position or the 
momentum operator by more appropriate operators that are unaffected by such 
subtleties.'' Starting from the original position and momentum operators 
$\hat x$ and $\hat p = - i \p_x$, which no longer obey the canonical 
commutation relation on the half-line, we were led to the operators 
$\hat x_R = \sigma_1 x$ and $\hat p_R = - i \sigma_1 \p_x$, which indeed satisfy
$[\hat x_R,\hat p_R] = i$. Moving on to an interval, we now realize that this 
relation again seizes to hold, which motivated us to replace $\hat x_R$ by the
more appropriate operator 
$\widetilde V_{\pi/L} = \exp(i \tfrac{\pi}{L} \hat x_R)$ that obeys the 
commutation relations eq.(\ref{pVcommutators}).

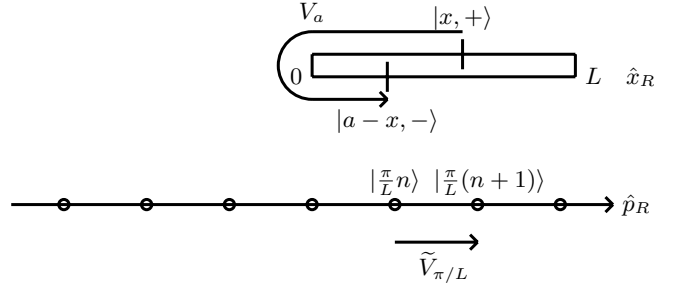
\begin{figure}[tbp]
\vspace{3cm}
\begin{tikzpicture}
\tikz\draw
[line width=0.4mm,color=black] (4,0) -- (7.5,0)
[line width=0.4mm,color=black] (4,0) -- (4,-0.3)
node[left] {$0$}
[line width=0.4mm,color=black] (4,-0.3) -- (7.5,-0.3)
[line width=0.4mm,color=black] (7.5,0) -- (7.5,-0.3)
node[right] {$L \quad \hat x_R$}
[line width=0.4mm,color=black] (6,-0.2) -- (6,0.2)
node[above] {$|x,+\rangle$}
[line width=0.4mm,color=black] (5,-0.5) -- (5,-0.1)
[line width=0.4mm,color=black] (6,0.3) -- (4,0.3)
node[above] {$V_a$}
[line width=0.4mm,color=black] (4,-0.6) -- (5,-0.6)
node[below] {$|a - x,-\rangle$}
[line width=0.4mm,color=black] (4,0.3) arc (90:270:0.45)
[line width=0.4mm,color=black] (4.9,-0.5) -- (5,-0.6)
[line width=0.4mm,color=black] (4.9,-0.7) -- (5,-0.6)
[line width=0.4mm,color=black] (0,-2) -- (8,-2)
node[right] {$\hat p_R$}
[line width=0.4mm,color=black] (7.9,-1.9) -- (8,-2)
[line width=0.4mm,color=black] (7.9,-2.1) -- (8,-2)
[line width=0.4mm,color=black] (0.7,-2) circle (0.5ex)
[line width=0.4mm,color=black] (1.8,-2) circle (0.5ex)
[line width=0.4mm,color=black] (2.9,-2) circle (0.5ex)
[line width=0.4mm,color=black] (4,-2) circle (0.5ex)
[line width=0.4mm,color=black] (5.1,-2) circle (0.5ex) 
node[above] {$|\tfrac{\pi}{L} n\rangle$}
[line width=0.4mm,color=black] (6.2,-2) circle (0.5ex)
node[above] {$\quad |\tfrac{\pi}{L} (n+1)\rangle$}
[line width=0.4mm,color=black] (7.3,-2) circle (0.5ex)
[line width=0.4mm,color=black] (5.1,-2.5) -- (6.2,-2.5)
node[below] {\hspace{-1cm} $\widetilde V_{\pi/L}$}
[line width=0.4mm,color=black] (6.1,-2.4) -- (6.2,-2.5)
[line width=0.4mm,color=black] (6.1,-2.6) -- (6.2,-2.5);
\end{tikzpicture}
\vspace{-3.5cm}
\caption{\it Action of the translation operators $V_a = \exp(i \hat p_R a)$ and
$\widetilde V_{\pi/L} = \exp(i \tfrac{\pi}{L} \hat x_R)$ in the continuous
position and discrete momentum eigenstates for the interval $[0,L]$.}
\label{Fig3}
\end{figure}

How does the operator $V_a = \exp(i \hat p_R a)$ act on the  $\hat x_R$
eigenstates $|x,\pm \rangle$ (with $x \in [0,L]$) of eq.(\ref{xReigenstates})? 
In close analogy to eq.(\ref{positionshift}), for $a \in [- L,L)$ one obtains
\begin{eqnarray}
&&V_a |x,+\rangle = \sigma_L |a - x + 2 L,-\rangle \ , \ 
\mbox{for} \ L \leq x - a < 2 L \ , \nonumber \\
&&V_a |x,+\rangle = |x - a,+\rangle \ , \ 
\mbox{for} \ 0 \leq x - a < L \ , \nonumber \\
&&V_a |x,+\rangle = \sigma |a - x,-\rangle \ , \ 
\mbox{for} \ - L \leq x - a < 0 \ , \nonumber \\
&&V_a |x,-\rangle = \sigma_L^* |- x - a + 2 L,+\rangle \ , \ 
\mbox{for} \ L \leq x + a < 2 L \ , \nonumber \\
&&V_a |x,-\rangle = |x + a,-\rangle \ , \ 
\mbox{for} \ 0 \leq x + a < L \ , \nonumber \\
&&V_a |x,-\rangle = \sigma^* |- x - a,+\rangle \ , \ 
\mbox{for} \ - L \leq x + a < 0 \ .
\label{xRshifts}
\end{eqnarray}
Independent of $\sigma$ or $\sigma_L$, as a consequence of 
eq.(\ref{pVcommutators}) one then obtains the Weyl group relation
\begin{equation}
V_a \widetilde V_{\pi/L} = 
\exp\left(i \frac{\pi}{L} a\right) \widetilde V_{\pi/L} V_a \ .
\label{WeylxRV}
\end{equation}
The action of the operators $V_a$ and $\widetilde V_{\pi/L}$ is illustrated in 
Fig.\ref{Fig3}. Since in the interval momentum is quantized, in the 
corresponding momentum space there are only discrete translations by multiples 
of $\tfrac{\pi}{L}$, which are generated by $\widetilde V_{\pi/L}$. Despite the 
fact that there is no translation symmetry in the interval $[0,L]$ (which is 
the reason why the original momentum operator $\hat p = - i \p_x$ is not 
self-adjoint), the momentum operator $\hat p_R = - i \sigma_1 \p_x$ indeed 
generates infinitesimal translations in the doubly covered interval $[0,L]^2$. 
As illustrated in Fig.\ref{Fig3}, $V_a = \exp(i \hat p_R a)$ acts like a 
periodic ``conveyor belt'', which transports states $|x,+\rangle$ (in the 
finite-energy sector) to the left, and states $|x,-\rangle$ (which lie in the
sector of states whose energy is ultraviolet-sensitive) to the right. When a 
state $|x,+\rangle$ is transported beyond the origin (at $x = 0$), it turns 
into the state $|a - x,-\rangle$. When such a state is transported further 
beyond the other end of the interval (at $x = L$), it returns to the 
finite-energy sector.

It is a matter of definition whether one wants to classify the situation in the
interval as belonging to ``canonical quantization''. We prefer to do so,
despite the fact that $\hat x_R$ and $\hat p_R$ no longer satisfy a meaningful
canonical commutation relation. The reason for this is simply that in a finite
volume momentum is quantized. As a consequence, $\hat x_R$ can no longer 
generate infinitesimal translations in momentum space. The appropriate discrete 
momentum translations by $\tfrac{\pi}{L}$ are generated by 
$\widetilde V_{\pi/L} = \exp(i \tfrac{\pi}{L} \hat x_R)$. As a result, the
commutation relations of eq.(\ref{pVcommutators}) replace the usual canonical 
commutation relation. In any case, the corresponding Weyl group of
eq.(\ref{WeylxRV}) has the same form as in the other cases of canonical
quantization, and thus using the term ``canonical quantization'' also for the
interval is indeed justified.

\subsection{Comparison with motion on $S^1$}

Let us compare the situation in an interval with the well-understood motion of 
a quantum mechanical particle on a circle $S^1$, parametrized by the angle 
$\varphi \in [-\pi,\pi]$. In that case, the operator for the linear momentum is 
replaced by the angular momentum operator $\hat L  = - i \p_\varphi$. First, let 
us investigate the Hermiticity and self-adjointness of $\hat L$, which acts on 
wave functions $\Psi(\varphi)$. Applying partial integration one obtains
\begin{eqnarray}
&&\langle \hat L^\dagger \chi|\Psi\rangle = \langle\chi|\hat L \Psi\rangle =
\nonumber \\
&&\langle \hat L \chi|\Psi\rangle -
i [\chi(\pi)^* \Psi(\pi) \! - \! \chi(-\pi )^* \Psi(- \pi)].
\label{LHermiticity}
\end{eqnarray}
Hermiticity of $\hat L$ hence requires that the expression in square brackets
vanishes. Since, in contrast to an open interval, $\varphi = \pi$ and 
$\varphi = - \pi$ parametrize the same point on the closed circle, the boundary 
condition
\begin{equation}
\Psi(\pi) = \rho \Psi(- \pi) \ , 
\end{equation}
which restricts the domain $D(\hat L)$, is local and thus physically admissible.
Inserting this relation in the square bracket in eq.(\ref{LHermiticity}), and 
using the fact that $\Psi(- \pi)$ can still take arbitrary values, one obtains 
the Hermiticity condition
\begin{equation}
[\chi(\pi)^* \rho - \chi(- \pi)^*] \Psi(- \pi) = 0 \ \Rightarrow \  
\chi(\pi) = \frac{1}{\rho^*} \chi(- \pi) \ ,
\end{equation}
which defines the domain $D(\hat L^\dagger)$. The operator $\hat L$ is 
self-adjoint only if $D(\hat L) = D(\hat L^\dagger)$, which is the case when
\begin{equation}
\rho = \frac{1}{\rho^*} \ \Rightarrow \ \rho = \exp(i \theta) \ .
\end{equation}
It is well-known that the angular momentum operator $\hat L$ has a 1-parameter
family of self-adjoint extensions parametrized by the angle $\theta$. The 
angular momentum eigenstates $|n\rangle$ then obey
\begin{eqnarray}
&&\hat L|n\rangle = \left(n + \frac{\theta}{2 \pi}\right) |n\rangle \ , \quad 
n \in \Z \ , \nonumber \\
&&\langle \varphi|n\rangle = \frac{1}{\sqrt{2 \pi}} 
\exp\left(i \left(n + \frac{\theta}{2 \pi}\right) \varphi\right) \ , 
\nonumber \\
&&\langle \pi|n\rangle = \exp(i \theta) \langle - \pi|n\rangle \ .
\end{eqnarray}

It is also well-known that $\hat \varphi$ and $\hat L$ do not obey the standard
canonical commutation relation. This is because an operation with the 
non-periodic operator $\hat \varphi$ leads out of the domain of $D(\hat L)$. 
Instead, the periodic unitary operator $\widetilde U = \exp(i \hat \varphi)$ 
acts as a discrete translation operator of angular momentum, with which it 
obeys the commutation relations
\begin{equation}
[\hat L,\widetilde U] = \widetilde U \ , \quad
[\hat L,\widetilde U^\dagger] = - \widetilde U^\dagger \ .
\label{LUcommutators}
\end{equation} 

Defining angular eigenstates $|\varphi\rangle$ by 
$\hat \varphi |\varphi\rangle = \varphi |\varphi\rangle$, the operator 
$U_\alpha = \exp(i \hat L \alpha)$, with $\alpha \in [- \pi,\pi)$, acts as
\begin{eqnarray}
&&U_\alpha|\varphi\rangle = \exp(i \theta) |\varphi - \alpha + 2 \pi\rangle, 
\mbox{for} - 2 \pi \leq \varphi - \alpha < - \pi, \nonumber \\  
&&U_\alpha|\varphi\rangle = |\varphi - \alpha\rangle, 
\mbox{for} - \pi \leq \varphi - \alpha < \pi, \nonumber \\  
&&U_\alpha|\varphi\rangle = \exp(- i \theta) |\varphi - \alpha - 2 \pi\rangle, 
\mbox{for} \ \pi \leq \varphi - \alpha < 2 \pi.
\end{eqnarray}
As a result of eq.(\ref{LUcommutators}), one then obtains the Weyl group 
relation
\begin{equation}
U_\alpha \widetilde U = \exp(i \alpha) \widetilde U U_\alpha \ .
\end{equation}
\begin{figure}[tbp]
\vspace{7cm}
\begin{tikzpicture}
\tikz\draw
[line width=0.4mm,color=black] (4,-3) circle (2)
[line width=0.4mm,color=black] (4+2.5/1.414,-3+2.5/1.414) arc (45:-45:2.5)
[line width=0.4mm,color=black] (4+2.5/1.414,-3-2.5/1.414) -- 
(4+2.5/1.414+0.1*1.414,-3-2.5/1.414)
[line width=0.4mm,color=black] (4+2.5/1.414,-3-2.5/1.414) -- 
(4+2.5/1.414,-3-2.5/1.414+0.1*1.414)
node[right] {$\quad U_\alpha$}
[line width=0.4mm,color=black] (5.9,-3.1) -- (6,-3)
node[left] {$\hat \varphi$}
[line width=0.4mm,color=black] (6.1,-3.1) -- (6,-3)
[line width=0.4mm,color=black] (4+2.2/1.414,-3+2.2/1.414) -- 
(4+1.8/1.414,-3+1.8/1.414)
node[left] {$|\varphi\rangle$}
[line width=0.4mm,color=black] (4+2.2/1.414,-3-2.2/1.414) -- 
(4+1.8/1.414,-3-1.8/1.414)
node[left] {$|\varphi - \alpha\rangle$}
[line width=0.4mm,color=black] (0,-6) -- (8,-6)
node[right] {$\hat L$}
[line width=0.4mm,color=black] (7.9,-5.9) -- (8,-6)
[line width=0.4mm,color=black] (7.9,-6.1) -- (8,-6)
[line width=0.4mm,color=black] (0.7,-6) circle (0.5ex)
[line width=0.4mm,color=black] (1.8,-6) circle (0.5ex)
[line width=0.4mm,color=black] (2.9,-6) circle (0.5ex)
[line width=0.4mm,color=black] (4,-6) circle (0.5ex)
[line width=0.4mm,color=black] (5.1,-6) circle (0.5ex) 
node[above] {$|n\rangle$}
[line width=0.4mm,color=black] (6.2,-6) circle (0.5ex)
node[above] {$|n+1\rangle$}
[line width=0.4mm,color=black] (7.3,-6) circle (0.5ex)
[line width=0.4mm,color=black] (5.1,-6.5) -- (6.2,-6.5)
node[below] {\hspace{-1cm} $\widetilde U$}
[line width=0.4mm,color=black] (6.1,-6.4) -- (6.2,-6.5)
[line width=0.4mm,color=black] (6.1,-6.6) -- (6.2,-6.5);
\end{tikzpicture}
\vspace{-7.5cm}
\caption{\it Action of the translation operators 
$U_\alpha = \exp(i \hat L \alpha)$ and
$\widetilde U = \exp(i \hat \varphi)$ on the continuous angle and discrete 
angular momentum eigenstates for the circle $S^1 = [- \pi,\pi]$.}
\label{Fig4}
\end{figure}
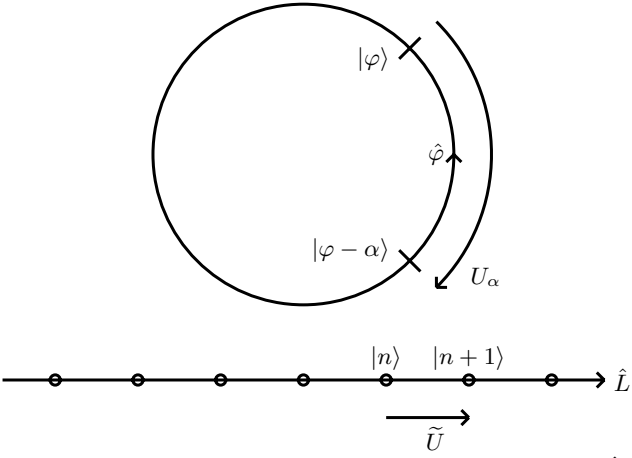
The action of the operators $U_\alpha$ and $\widetilde U$ is illustrated in 
Fig.\ref{Fig4}. Angular momentum is quantized in integer units (shifted by 
$\frac{\theta}{2 \pi}$), and the corresponding discrete translation symmetry is 
generated by $\widetilde U$. The continuous angular rotations described by
$U_\alpha = \exp(i \hat L \alpha)$ act like the periodic ``conveyor belt'' 
associated with $V_a = \exp(i \hat p_R a)$ for the interval that is illustrated 
in Fig.\ref{Fig3}.

The similarities between eq.(\ref{LUcommutators}) for the circle 
$S^1 = [- \pi,\pi]$ and eq.(\ref{pVcommutators}) for the interval $[0,L]$ are 
not accidental. In fact, the operator pairs $\hat L$, $\widetilde V$ and 
$\hat p_R$, $\widetilde V_{\pi/L}$ are mathematically related by a 
unitary transformation, although they are physically quite different. Up to
phase factors, the unitary transformation $U$ maps the position eigenstates 
$|x,\pm\rangle$ in the interval $x \in [0,L]$ to the angular eigenstates
$|\varphi\rangle$ on the circle $\varphi \in [- \pi,\pi]$
\begin{eqnarray}
&&U V_a U^\dagger = U_{\alpha = \pi a/L} \ , \quad 
U \widetilde V_{\pi/L} U^\dagger = \widetilde U \ , \nonumber \\
&&U \hat x_R U^\dagger = \frac{L}{\pi} \hat \varphi \ , \nonumber \\ 
&&U |x,+\rangle = |\varphi = \tfrac{\pi x}{L}\rangle \ , \quad
U |x,-\rangle = \sigma^* |\varphi = - \tfrac{\pi x}{L}\rangle \ , \nonumber \\
&&U \hat p_R U^\dagger = \frac{\pi}{L} \hat L \ , \quad 
U|\phi_k\rangle = |n = \tfrac{k L}{\pi}\rangle \ .
\label{Uintervalcircle}
\end{eqnarray}
Let us check this for consistency. For example, using eq.(\ref{xRshifts}) for 
$L \leq x - a < 2 L$, which implies $\pi \leq \varphi - \alpha < 2 \pi$, we 
obtain
\begin{eqnarray}
&&V_a |x,+\rangle = \sigma_L |a - x + 2 L,-\rangle \ \Rightarrow 
\nonumber \\
&&U V_a U^\dagger U |x,+\rangle = \sigma_L U |a - x + 2 L,-\rangle \ 
\Rightarrow \nonumber \\
&&U_{\alpha = \pi a/L} |\varphi = \tfrac{\pi x}{L}\rangle = 
\sigma_L \sigma^* |\varphi = - \tfrac{\pi}{L} (a - x + 2 L)\rangle \ 
\Rightarrow \nonumber \\
&&\exp(- i \theta) |\varphi = \tfrac{\pi}{L}(x - a) - 2 \pi\rangle
\nonumber \\
&&= \sigma_L \sigma^* 
|\varphi = \tfrac{\pi}{L}(x - a) - 2 \pi\rangle \ .
\end{eqnarray}
This is indeed consistent, because based on eq.(\ref{thetaequation}), 
$\exp(i \theta) = \sigma \sigma_L^*$.

Eq.(\ref{Uintervalcircle}) is very similar to eq.(\ref{Uhalflineaxis}) which
describes the mathematical unitary equivalence of the standard operators 
$\hat x$ and $\hat p$ acting on wave functions over the entire real axis with 
the operators $\hat x_R$ and $\hat p_R$ that apply on the half-line, that
followed from the Stone-von-Neumann theorem. Also in that case the mathematical 
unitary equivalence relates two physically very different situations.

\subsection{Physical significance of $\theta$}

For circular motion, the parameter $\theta$ is well-known to represent a 
magnetic flux that threads the circle, which affects a charged particle moving 
around the circle via an Aharonov-Bohm phase. First of all, since the angles 
$\varphi = \pm \pi$ parametrize one and the same point on the circle, in the
absence of magnetic flux the value of the wave function should be unique at
that point, i.e.\ $\Psi(\pi) = \Psi(- \pi)$, such that $\theta = 0$. In the
presence of a magnetic flux $\Phi$, the angular momentum operator takes the form
\begin{equation}
\hat L' = - i \p_\varphi + \frac{e \Phi}{2 \pi} = 
- i \p_\varphi + \frac{\theta}{2 \pi} \ , \quad \theta = e \Phi \ .
\end{equation}
Here $- e$ is the charge of the particle and (along with $\hbar$) we have put
the velocity of light to 1. One can now perform the unitary transformation 
\begin{eqnarray}
&&U(\theta) = \exp\left(i \frac{\theta}{2 \pi} \hat \varphi\right) \ ,
\nonumber \\
&&\hat L = U(\theta) \hat L' U(\theta)^\dagger = - i \p_\varphi \ , \
\Psi(\varphi) = U(\theta) \Psi'(\varphi) \ ,
\end{eqnarray}
which implies
\begin{eqnarray}
\Psi(\pi)&=&\exp\left(i \frac{\theta}{2}\right) \Psi'(\pi) = 
\exp\left(i \frac{\theta}{2}\right) \Psi'(- \pi) \nonumber \\
&=&\exp(i \theta) \Psi(- \pi) \ . 
\end{eqnarray}
After the unitary transformation, the wave function $\Psi(\varphi)$ is
no longer single-valued at $\varphi = \pm \pi$. This is no problem, because
the unitary transformation corresponds to a non-periodic gauge transformation.
Like the magnetic flux $\Phi$, the parameter $\theta$ itself is gauge invariant.
However, the unitary transformation $U(\theta)$ moves $\theta$ from the 
operator $\hat L'$ to the (no longer strictly periodic) wave function 
$\Psi(\varphi)$.

The similarity with the situation in the interval suggests that a similar
interpretation exists for the corresponding parameter $\theta$, which
results from
\begin{equation}
\exp(i \theta) = \sigma \sigma_L^* = 
\frac{(1 - \lambda)(1 + \lambda_L)}{(1 + \lambda)(1 - \lambda_L)} \ .
\end{equation}
As we have seen, the unitary transformation $W = \exp(i \omega \sigma_1)$ of 
eq.(\ref{Wequation}) changes the values of the self-adjoint extension 
parameters, such that $\sigma' = \exp(- 2 i \omega) \sigma$, 
$\sigma_L' = \exp(- 2 i \omega) \sigma_L$, but leaves $\theta$ invariant 
(cf.\ eq.(\ref{invarianttheta})). Still, we can perform another $x$-dependent
unitary transformation, 
\begin{eqnarray}
W(\theta) = \exp\left(i \frac{\theta}{2 L} \sigma_1 \hat x\right) = 
\exp\left(i \frac{\theta}{2 L} \hat x_R\right) \ ,  
\end{eqnarray}
which implies $\sigma' = \sigma$ and $\sigma_L' = \exp(- i \theta) \sigma_L$ 
such that $\sigma' {\sigma_L'}^* = 1$. The transformation $W(\theta)$ again
represents a gauge transformation in Hilbert space, which now moves $\theta$
from the boundary conditions on the wave function to the momentum operator
\begin{equation}
\hat p_R' = W(\theta)^\dagger \hat p_R W(\theta) = 
- i \sigma_1 \p_x + \frac{\theta}{2 L} \ .
\end{equation}

When we perform the unitary transformation on the kinetic energy $\hat T$ we
obtain
\begin{eqnarray}
&&\hat T' = W(\theta)^\dagger \hat T W(\theta) = \frac{{\hat p'}{^2}}{2 m} \1 \ ,
\nonumber \\
&&\hat p' = - i \p_x + \frac{\theta}{2 L} = \hat p + e A_x \ , \quad
\theta = 2 e A_x L \ . 
\end{eqnarray}
In this case, $\theta$ manifests itself as a vector potential $A_x$, or more
precisely as its line integral $\int_0^L dx A_x = A_x L$ that connects the 
two boundaries. The original Robin boundary condition, 
$\gamma \Psi(0) - \p_x \Psi(0) = 0$, of eq.(\ref{Robinbc}), along with its
counterpart at the other end of the interval, 
$\gamma_L \Psi(L) + \p_x \Psi(L) = 0$, must also be transformed accordingly, 
and one obtains \cite{AlH12}
\begin{eqnarray}
&&\gamma \Psi'(0) - D_x \Psi'(0) = 0 \ , \ 
\gamma_L \Psi'(L) + D_x \Psi'(L) = 0 \ , \nonumber \\
&&D_x = \p_x + i e A_x \ .
\label{Robinbctransformed}
\end{eqnarray}
The covariant derivative $D_x$ also enters the conserved probability current
\begin{equation}
j'(x) = 
\frac{1}{2 m i}\left(\Psi'(x)^* D_x \Psi'(x) - D_x \Psi'(x)^* \Psi'(x)\right) 
\ .
\end{equation}
As a result, the boundary conditions of eq.(\ref{Robinbctransformed}) still
guarantee that no probability leaks out of the interval, i.e.\ 
$j'(0) = j'(L) = 0$.

Under a general gauge transformation $\varphi(x)$ the vector potential and the
wave function transform as
\begin{equation}
^\varphi A_x(x) = A_x(x) - \p_x \varphi(x),
^\varphi\Psi'(x) = \exp\left(i e \varphi(x)\right) \Psi'(x) \ .
\end{equation}
The gauge string connecting the two ends of the interval, capped by the values 
of the wave function at the end points,
\begin{eqnarray}
&&\!\!\!\!S = 
|0\rangle \exp\left(i e \int_0^L dx \ A_x(x)\right) \langle L| 
\ , \nonumber \\
&&\!\!\!\!\langle \Psi'|S|\Psi'\rangle =
\Psi'(0)^* \exp\left(i e \int_0^L dx \ A_x(x)\right) \Psi'(L),
\end{eqnarray}
is gauge invariant, i.e.\ $\langle ^\varphi\Psi'|^\varphi S|^\varphi\Psi'\rangle = 
\langle\Psi'|S|\Psi'\rangle$. In cases where such a gauge string stretches 
through the interval, the parameter $\theta$ appears in the corresponding 
momentum operator.

\subsection{Momentum measurements in $[0,L]$}

Let us now consider momentum measurements in the interval $[0,L]$. For
simplicity, we consider a Hamiltonian without a potential (i.e.\ $V(x) = 0$).
First, we investigate Neumann boundary conditions, which are characterized by 
$\gamma = 0$ at both ends of the interval. In the limit 
$\mu \rightarrow \infty$, the finite-energy eigenstates of $\hat H(\mu)$ then 
take the form
\begin{eqnarray}
&&\hat H(\mu) \psi_l(x) = E_l \psi_l(x) \ , \quad 
E_l = \frac{\pi^2 l^2}{2 m L^2} \ , \quad l \in \N_{\geq 0} \ , \nonumber \\ 
&&\psi_0(x) =  
\frac{1}{\sqrt{2 L}} \left(\begin{array}{c} 1 \\ 1 \end{array}\right) \ ,
\nonumber \\ 
&&\psi_{l > 0}(x) = \frac{1}{\sqrt{L}}
\left(\begin{array}{c} \cos(\pi l x/L) \\ \cos(\pi l x/L)
\end{array}\right) \ .
\end{eqnarray}
Again for simplicity, we choose $\lambda_L = \lambda$ which implies 
$\theta = 0$, such that the corresponding momentum eigenvalues and 
eigenfunctions are
\begin{eqnarray}
&&\hat p_R \phi_k(x) = k \phi_k(x), \ k = \frac{\pi n}{L}, \ n \in \Z, \ 
\sigma = \frac{1 - \lambda}{1 + \lambda} \in U(1) \ , \nonumber \\ 
&&\phi_k(x) = \frac{1}{2 \sqrt{L}}
\left(\begin{array}{c} \exp(i k x) + \sigma \exp(- i k x) 
\\ \exp(i k x) - \sigma \exp(- i k x) \end{array}\right) \ .
\end{eqnarray}

\begin{figure}[thb]
\includegraphics[width=0.48\textwidth]{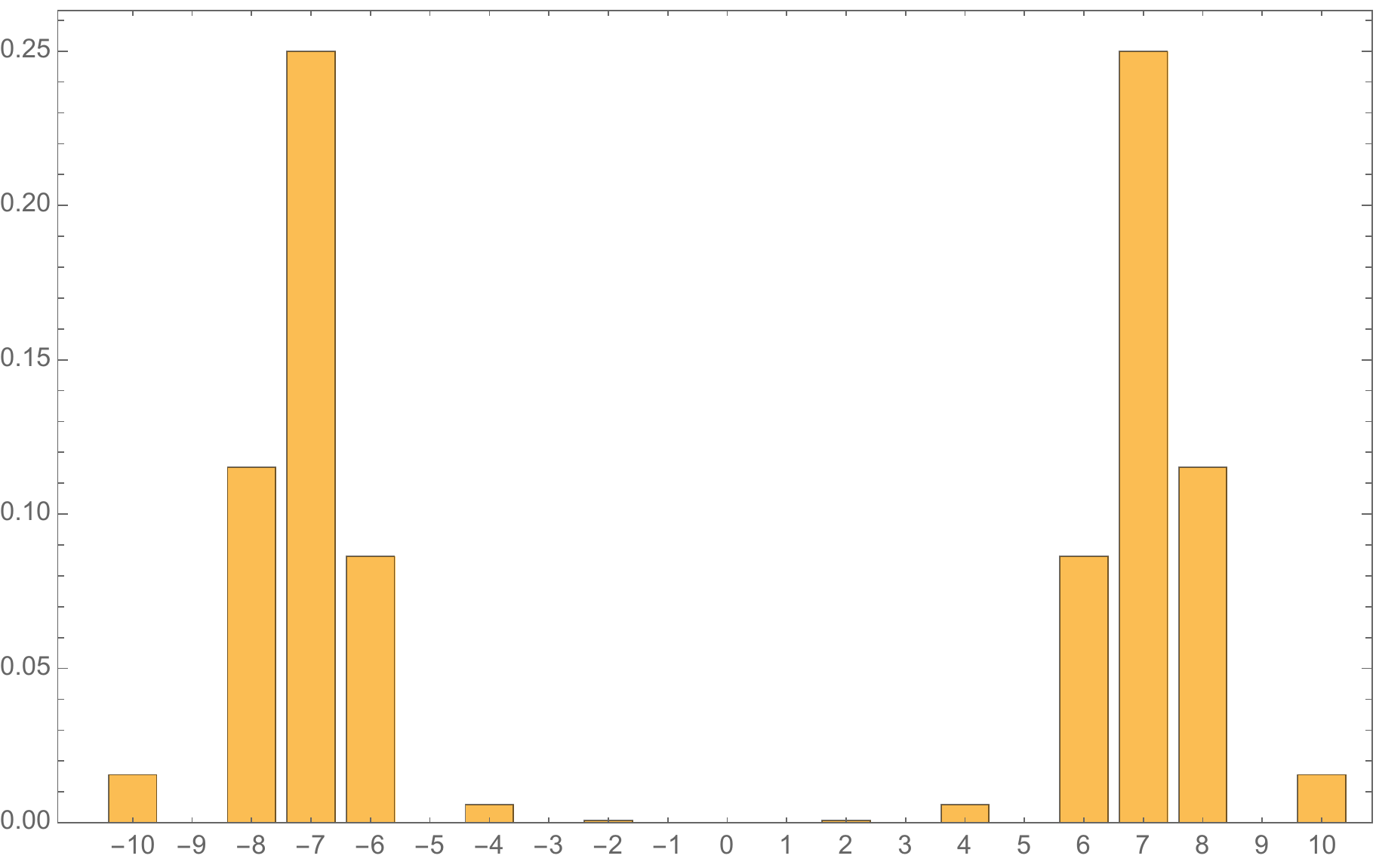}
\caption{\it Probability to measure the momentum $k = \tfrac{\pi}{L} n$ in the
energy eigenstate $\psi_l(x)$ with $l = 7$ for Neumann boundary conditions, as
a function of $n \in \{-10,\dots,10\}$.}
\label{plot1}
\end{figure}

First of all, one obtains $\langle \psi_l|\hat p_R|\psi_l\rangle = 0$. 
Irrespective of the value of $\lambda$, when one projects them onto the 
finite-energy sector, the momentum eigenstates are just 
$\phi^+_{k,e}(x) = \phi^+_{k,o}(x) = \frac{1}{2 \sqrt{L}} \exp(i k x)$. In the 
ground state the probability to measure the momentum value $k = 0$ is
$|\langle \phi_0|\psi_0\rangle|^2 = \tfrac{1}{2}$. Similarly, the probability
to measure $k = \tfrac{\pi n}{L}$ with $n \neq 0$ is 
\begin{equation}
|\langle \phi_k|\psi_0\rangle|^2 = \frac{2}{\pi^2 n^2} \ ,
\end{equation}
for odd $n$ and zero otherwise. When one measures the momentum in an energy 
eigenstate $\psi_l(x)$ with $l > 0$, one obtains $k = \pm \tfrac{\pi l}{L}$ each
with probability $\tfrac{1}{4}$. The probability to measure 
$k = \tfrac{\pi n}{L}$ for $n \neq \pm l$ is then given by
\begin{equation} 
|\langle \phi_k|\psi_l\rangle|^2 = \frac{4 n^2}{\pi^2 (l^2 - n^2)^2} \ ,
\label{phikpsiloverlap}
\end{equation} 
if $(-1)^n = - (-1)^l$ and zero otherwise. Indeed the various probabilities,
which are illustrated in Fig.\ref{plot1}, are correctly normalized because
\begin{eqnarray}
&&\frac{1}{2} + \!\! \sum_{n \in \Z, n \, \tiny{\mbox{odd}}} 
\frac{2}{\pi^2 n^2} = 1 \ , \ \mbox{for $l = 0$} \ , \nonumber \\
&&\frac{1}{4} + \frac{1}{4} + \!\! \sum_{n \in \Z, n \, \tiny{\mbox{even}}} 
\frac{4 n^2}{\pi^2 (l^2 - n^2)^2} = 1, \mbox{for odd $l > 0$},
\nonumber \\
&&\frac{1}{4} + \frac{1}{4} + \!\! \sum_{n \in \Z, n \, \tiny{\mbox{odd}}} 
\frac{4 n^2}{\pi^2 (l^2 - n^2)^2} = 1, \mbox{for even $l > 0$}.
\end{eqnarray}

Let us also consider the momentum uncertainty 
$(\Delta p_R)^2 = \langle \hat p_R^2\rangle - \langle \hat p_R\rangle^2$ in
the energy eigenstate $\psi_l(x)$. Besides $\langle \hat p_R\rangle = 0$ one 
obtains
\begin{eqnarray}
\langle \hat p_R^2\rangle&=&\frac{\pi^2}{L^2} 
\sum_{n \in \Z, n \, \tiny{\mbox{odd}}} \frac{2}{\pi^2} \rightarrow \infty \ , \ 
\mbox{for $l = 0$} \ , \nonumber \\
\langle \hat p_R^2\rangle&=&\frac{\pi^2}{L^2} 
\left(\frac{l^2}{4} + \frac{(- l)^2}{4} + \sum_{n \in \Z, n \, \tiny{\mbox{even}}}
\frac{4 n^4}{\pi^2 (l^2 - n^2)^2}\right) \nonumber \\
&\rightarrow& \infty \ , \ \mbox{for odd $l > 0$} \ , \nonumber \\
\langle \hat p_R^2\rangle&=&\frac{\pi^2}{L^2} 
\left(\frac{l^2}{4} + \frac{(- l)^2}{4} + \sum_{n \in \Z, n \, \tiny{\mbox{odd}}} 
\frac{4 n^4}{\pi^2 (l^2 - n^2)^2}\right) \nonumber \\
&\rightarrow&\infty \ , \ \mbox{for even $l > 0$} \ .
\end{eqnarray}
Hence, for Neumann boundary conditions the momentum uncertainty diverges for 
any energy eigenstate. This is a consequence of the domain incompatibility
$D(\hat T) \neq D(\hat p_R)$. Although at a formal level of differential
expressions the kinetic energy operator $\hat T$ and the momentum operator 
$\hat p_R$ seem to commute (which would imply a vanishing momentum uncertainty),
the domain incompatibility leads to a completely different result.

Let us now consider the limit $L \rightarrow \infty$ in which we expect
to recover the results for the half-line. For $\gamma = 0$, according to
eq.(\ref{Requation}), $R(p) = 1$. Then, using eq.(\ref{kpsiEoverlap}), for
$k \neq \pm p$ one obtains
\begin{equation}
|\langle k|\psi_E\rangle|^2 = \frac{4 k^2}{(p^2 - k^2)^2} \ , \quad 
E= \frac{p^2}{2m} \ .
\end{equation}
For finite $L$ we identify $k = \tfrac{\pi n}{L}$ and $p = \tfrac{\pi l}{L}$,
such that, using eq.(\ref{phikpsiloverlap}) for $n \neq \pm l$, one gets 
\begin{eqnarray} 
|\langle \phi_k|\psi_l\rangle|^2&=&\frac{4 n^2}{\pi^2 (l^2 - n^2)^2} =
\frac{1}{L^2} \frac{4 k^2}{(p^2 - k^2)^2} \nonumber \\
&=&\frac{1}{L^2} |\langle k|\psi_E\rangle|^2 \ .
\end{eqnarray}
The factor $\tfrac{1}{L^2}$ is due to the fact that $|k\rangle$ and 
$|\Psi_E\rangle$ are normalized to $\delta$-functions, while $|\phi_k\rangle$
and $|\psi_l\rangle$ are normalized to 1. In any case, this implies that (with
Neumann boundary conditions) in the stationary scattering state 
$\psi_E(x) = \exp(- i p x) + \exp(i p x)$ the probability to measure a momentum
value $k = p$ or $k = - p$ is $\tfrac{1}{4}$ in both cases. In the remaining 
half of the cases the momentum measurement returns a result $k \neq \pm p$ with 
a divergent momentum uncertainty. This somewhat counter-intuitive result is 
again due to domain incompatibilities.

\begin{figure}[thb]
\includegraphics[width=0.48\textwidth]{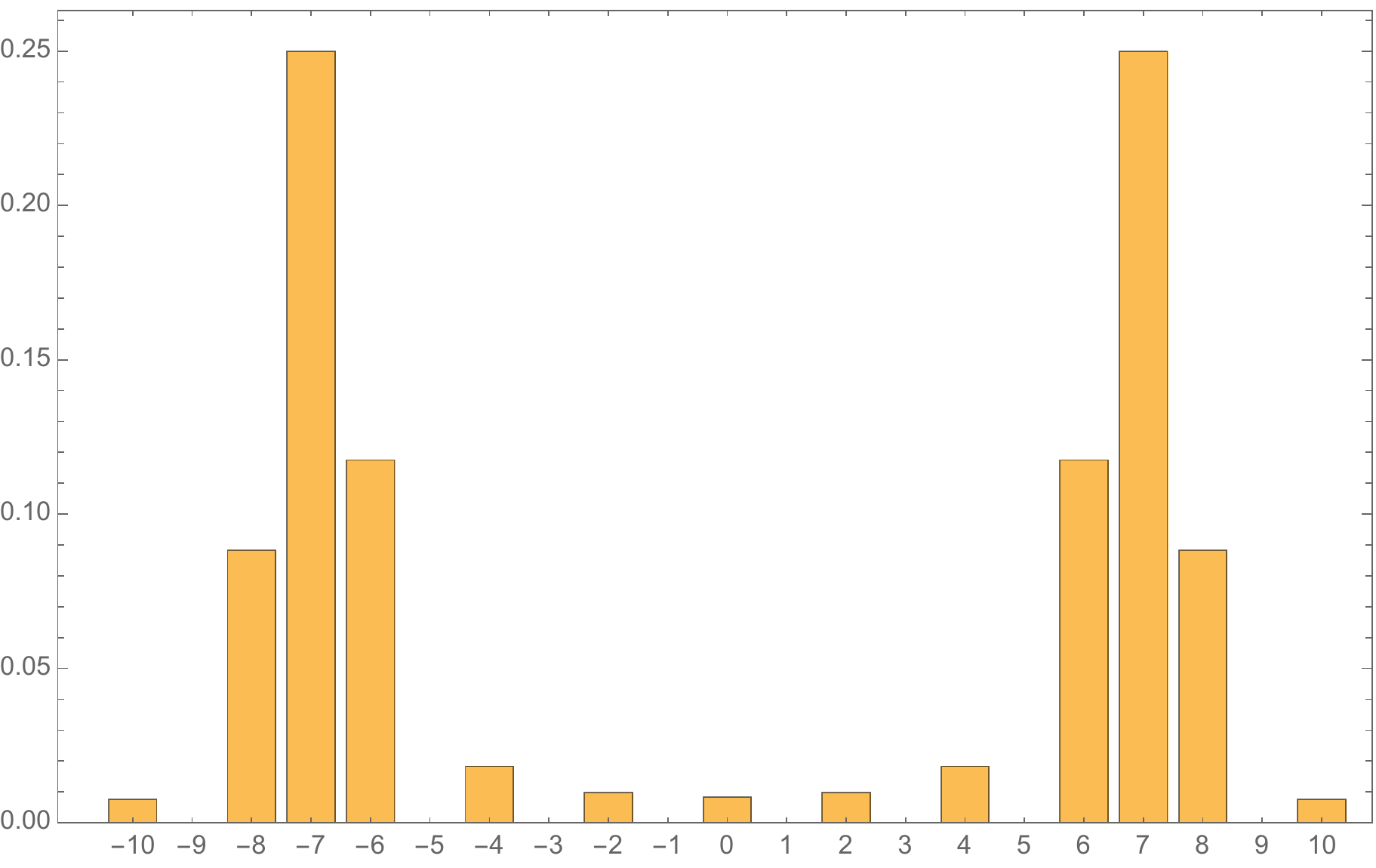}
\caption{\it Probability to measure the momentum $k = \tfrac{\pi}{L} n$ in the
energy eigenstate $\psi_l(x)$ with $l = 7$ for Dirichlet boundary conditions, 
as a function of $n \in \{-10,\dots,10\}$.}
\label{plot2}
\end{figure}

Finally, let us also discuss the standard textbook case of Dirichlet boundary 
conditions which are characterized by $\gamma = \infty$. The spectrum of
finite-energy states then takes the form
\begin{eqnarray}
&&\hat H(\mu) \psi_l(x) = E_l \psi_l(x) \ , \quad 
E_l = \frac{\pi^2 l^2}{2 m L^2} \ , \quad l \in \N_{> 0} \ , \nonumber \\ 
&&\psi_l(x) = \frac{1}{\sqrt{L}}
\left(\begin{array}{c} \sin(\pi l x/L) \\ \sin(\pi l x/L)
\end{array}\right) \ .
\end{eqnarray}
In this case, the probability to measure the momentum $k = p$ or $k = - p$ is
again $\tfrac{1}{4}$ in both cases, and the probability to measure 
$k \neq \pm p$ is
\begin{equation} 
|\langle \phi_k|\psi_l\rangle|^2 = \frac{4 l^2}{\pi^2 (l^2 - n^2)^2} \ ,
\end{equation}
if $(-1)^n = - (-1)^l$ and zero otherwise. Again, the probabilities, which are
illustrated in Fig.\ref{plot2}, are correctly normalized because
\begin{eqnarray}
&&\frac{1}{4} + \frac{1}{4} + \!\! \sum_{n \in \Z, n \, \tiny{\mbox{even}}} 
\frac{4 l^2}{\pi^2 (l^2 - n^2)^2} = 1, \mbox{for odd $l > 0$},
\nonumber \\
&&\frac{1}{4} + \frac{1}{4} + \!\! \sum_{n \in \Z, n \, \tiny{\mbox{odd}}} 
\frac{4 l^2}{\pi^2 (l^2 - n^2)^2} = 1, \mbox{for even $l > 0$}.
\end{eqnarray}
In this case one obtains
\begin{eqnarray}
\langle \hat p_R^2\rangle&=&\frac{\pi^2}{L^2} 
\left(\frac{l^2}{4} + \frac{(- l)^2}{4} + \sum_{n \in \Z, n \, \tiny{\mbox{even}}}
\frac{4 l^2 n^2}{\pi^2 (l^2 - n^2)^2}\right) \nonumber \\
&=&\frac{\pi^2 l^2}{L^2} \ , \ \mbox{for odd $l > 0$} \ , \nonumber \\
\langle \hat p_R^2\rangle&=&\frac{\pi^2}{L^2} 
\left(\frac{l^2}{4} + \frac{(- l)^2}{4} + \sum_{n \in \Z, n \, \tiny{\mbox{odd}}} 
\frac{4 l^2 n^2}{\pi^2 (l^2 - n^2)^2}\right) \nonumber \\
&=&\frac{\pi^2 l^2}{L^2} \ , \ \mbox{for even $l > 0$} \ ,
\end{eqnarray}
which implies $(\Delta p_R)^2 = 2 m E_l$. Hence, for Dirichlet boundary
conditions the momentum uncertainty in an energy eigenstate is finite. This is
because, in this case, $D(\hat T) \subset D(\hat p_R)$. Again using 
eq.(\ref{Requation}), one now gets $R(p) = - 1$, and following 
eq.(\ref{kpsiEoverlap}), for $k \neq \pm p$ one obtains
\begin{equation}
|\langle k|\psi_E\rangle|^2 = \frac{4 p^2}{(p^2 - k^2)^2} = 
L^2 |\langle \phi_k|\psi_l\rangle|^2 \ .
\end{equation}
For Dirichlet boundary conditions the stationary scattering state is
$\psi_E(x) = \exp(- i p x) - \exp(i p x)$. Then the probability to measure 
momentum $k = p$ or $k = - p$ is again $\tfrac{1}{4}$ in both cases. In the 
remaining half of the cases the momentum measurement results in $k \neq \pm p$, 
now with a finite momentum uncertainty.

\section{Conclusions}

We have introduced a new concept for a self-adjoint quantum mechanical momentum 
operator for an interval $[0,L]$ and for the half-line $\R_{\geq 0}$. The 
new concept arises naturally in the continuum limit of the lattice-regularized
problem. On the lattice one distinguishes even and odd lattice points. In
the continuum limit, this naturally leads to a two-component wave function, 
which is associated with a doubling of the Hilbert space from $L^2([0,L])$ to
$L^2([0,L]^2)$ and from $L^2(\R_{\geq 0})$ to $L^2(\R_{\geq 0}^2)$. The additional
continuum states correspond to lattice states with energies at the cut-off 
scale. In the continuum limit, these states are removed from the physical energy
spectrum. The key insight underlying the new concept is that these states, 
although they are ultimately removed to infinite energy, must be kept in the 
physical description in order to facilitate the construction of a self-adjoint 
momentum operator. Interestingly, although the resulting momentum operator for
the half-line is endowed with a self-adjoint extension parameter $\lambda$
(associated with the origin) and is thus not unique, the results of momentum 
measurements performed on finite-energy states are independent of this 
parameter. In an interval, the momentum operator is characterized by two 
self-adjoint extension parameters $\lambda$ and $\lambda_L$ (associated with 
the two boundary points). In that case, the value of the quantized momentum, 
$k = \frac{\pi}{L}(n + \tfrac{\theta}{2 \pi})$, depends on the particular 
combination $\theta$ of the two self-adjoint extension parameters.

Based upon the new concept for the momentum operator, canonical quantization 
becomes applicable both to the half-line and to an interval. However, due 
to the existence of sharp boundaries, self-adjoint extension parameters enter 
the description and thus lead to physically in-equivalent quantum variants of 
the same classical system. This goes beyond the usual operator ordering 
ambiguities. In particular, different operators, like the momentum $\hat p_R$ 
and the Hamiltonian $\hat H(\mu)$, act in different domains, 
$D(\hat p_R) \neq D(\hat H(\mu))$, of the Hilbert space. As a consequence, the 
commutation relations that result from classical Poisson bracket relations are 
just formal equations relating differential expressions. Understanding the true 
nature of the relations between the various operators requires a careful 
analysis of the corresponding operator domains. 

This is an inevitable consequence of the low-energy continuum description of a
system with sharp impenetrable boundaries, which are necessarily
ultraviolet sensitive. Working explicitly with an ultraviolet lattice cut-off
(representing the shortest physically relevant distance scale) is
straightforward and might even be quite physical, but is not very transparent. 
A transparent effective low-energy continuum description, as it is completely
common in quantum mechanics, necessarily uses an infinite-dimensional Hilbert 
space. As we have seen, on the half-line or in an interval this requires
a careful application of the canonical quantization procedure. Familiarizing 
ourselves with the elegant mathematical framework originally established by
von Neumann is very well worth the effort in order to gain a more complete 
physical understanding of these ``simple'' quantum systems.

It should be mentioned that most of the somewhat subtle Hilbert space and 
operator domain issues, associated with the canonical quantization procedure 
that we carried out above, arose only because we decided to construct the 
momentum operator in addition to just the Hamiltonian. To a physicist who favors
Everett's many-worlds interpretation of quantum mechanics \cite{Eve57}, this 
may seem unnecessary, because any measurement process is then incorporated in 
the global Hamiltonian that governs the time-evolution of the wave function of 
the entire universe. This wave function includes the quantum system under 
study, any device that registers measurement results, as well as the conscious 
observer who reads off those results and uses them to draw conclusions about 
how the quantum system works. Still, when engaging in an actual study of an 
isolated quantum system (rather than of the universe as a whole), even a 
hard-line ``Everettian'' would probably prefer to work with the much more 
tractable canonical quantization procedure applied above.

The new concept of momentum in a space with boundaries has potentially far
reaching consequences. Its generalization to higher dimensions is 
straightforward and was sketched in \cite{AlH20}. A natural next step, which 
is currently under investigation, concerns the generalization and physical 
interpretation of the Heisenberg uncertainty relation that was derived for a
finite volume with sharp boundaries in \cite{AlH12}. This can be applied, for
example, to optical box traps \cite{Gau13} and quantum dots \cite{Har05}, which 
may make the new momentum concept experimentally accessible. In this context,
it is also interesting to construct a momentum measurement device, at least at
a theoretical level, for example, along the lines originally introduced by
von Neumann \cite{Neu32a}. This implies to couple the quantum particle to 
another quantum system that serves as a measurement device, whose pointer can 
then be read out at the classical level. Time-of-flight momentum measurements 
of this kind have been discussed, for example, in \cite{DiP19}. Further 
applications, again along the lines of \cite{AlH12}, suggest themselves in the 
context of relativistic fermions, for example, to the phenomenological MIT bag 
model \cite{Cho74,Cho74a,Has78}, or to domain wall fermions residing in an 
interval of extra-dimensional space \cite{Kap92,Sha93}. Canonical quantization 
(which was thought to be inapplicable because the standard momentum operator is 
not self-adjoint) can be applied in all these situations based on the new 
concept of momentum in a space with sharp boundaries.

\section{Acknowledgments}

We thank Matthias Blau for illuminating discussions.

\appendix

\section{Lattice regularization of the momentum 
operator}

\begin{figure}[thb]
\vspace{1.2cm}
\begin{tikzpicture}
\tikz\draw[line width=0.4mm,color=black] (0,0) -- (8.1,0)
[line width=0.4mm,color=black] (0,-0.3) -- (0,0.3)
[line width=0.4mm,color=black] (8.1,-0.3) -- (8.1,0.3)
[fill=black] (0.45,0) circle (0.5ex)
[fill=black] (1.35,0) circle (0.5ex)
[fill=black] (2.25,0) circle (0.5ex)
[fill=black] (3.15,0) circle (0.5ex)
[fill=black] (4.05,0) circle (0.5ex)
[fill=black] (4.95,0) circle (0.5ex)
[fill=black] (5.85,0) circle (0.5ex)
[fill=black] (6.75,0) circle (0.5ex)
[fill=black] (7.65,0) circle (0.5ex)
[line width=0.4mm,color=black] (0,-0.5) -- (0.45,-0.5)
node[below] {\hspace{-0.45cm}$a/2$}
[line width=0.4mm,color=black] (0,-0.6) -- (0,-0.4)
[line width=0.4mm,color=black] (0.45,-0.6) -- (0.45,-0.4)
[line width=0.4mm,color=black] (4.05,-0.5) -- (4.95,-0.5) 
node[below] {\hspace{-0.9cm}$a$}
[line width=0.4mm,color=black] (4.05,-0.6) -- (4.05,-0.4)
[line width=0.4mm,color=black] (4.95,-0.6) -- (4.95,-0.4);
\end{tikzpicture}
\vspace{-1.5cm}
\caption{\it Lattice with $N = 9$ points in the interval $[0,L]$.}
\label{Fig5}
\end{figure}
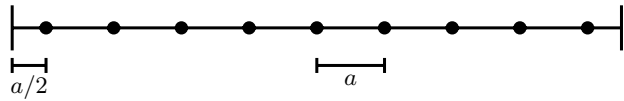

In order to circumvent the subtleties associated with Hermiticity versus 
self-adjointness, which arise because the Hilbert space is infinite-dimensional,
in \cite{AlH20} we have investigated the problem on the interval in an
ultraviolet lattice regularization. As illustrated in Fig.\ref{Fig5}, the 
interval $[0,L]$ is then divided into $N = L/a$ segments of size $a$ (not to be
confused with the self-adjoint extension parameter of eq.(\ref{generalbc})), 
with a lattice point in the middle of each segment, such that 
$x = (n - \tfrac{1}{2}) a$, $n \in \{1,2,\dots,N\}$. The Hilbert space then 
becomes $N$-dimensional and self-adjointness becomes indistinguishable from 
Hermiticity. As a result, there are no longer any domain issues, since every 
operator can act in the entire finite-dimensional Hilbert space.

The lattice momentum operator is represented by forward and backward
discretized derivatives. 
\begin{eqnarray}
&&\hat p_F = - \frac{i}{a} \left(\begin{array}{ccccccc}
-1 & 1 & 0 & \dots & 0 & 0 & 0 \\
 0 &-1 & 1 & \dots & 0 & 0 & 0 \\
 0 & 0 &-1 & \dots & 0 & 0 & 0 \\
 . & . & . & \dots & . & . & . \\
 . & . & . & \dots & . & . & . \\
 0 & 0 & 0 & \dots &-1 & 1 & 0 \\
 0 & 0 & 0 & \dots & 0 &-1 & 1 \\
 0 & 0 & 0 & \dots & 0 & 0 &\lambda_L \end{array}\right) \ , \nonumber \\
&&\hat p_B = - \frac{i}{a} \left(\begin{array}{ccccccc}
-\lambda& 0 & 0 & \dots & 0 & 0 & 0 \\
-1 & 1 & 0 & \dots & 0 & 0 & 0 \\
 0 &-1 & 1 & \dots & 0 & 0 & 0 \\
 . & . & . & \dots & . & . & . \\
 . & . & . & \dots & . & . & . \\
 0 & 0 & 0 & \dots & 1 & 0 & 0 \\
 0 & 0 & 0 & \dots &-1 & 1 & 0 \\
 0 & 0 & 0 & \dots & 0 &-1 & 1 \end{array}\right) \ .
\end{eqnarray}
On the lattice, the parameters $\lambda, \lambda_L \in i \R$ are directly
incorporated in the corresponding matrices. In the continuum limit 
$a \rightarrow 0$ they turn into self-adjoint extension parameters.
Neither $\hat p_F$ nor $\hat p_B$ are Hermitean matrices. It is natural to
construct the following combinations
\begin{eqnarray}
\hat p_R&=&\frac{1}{4}(\hat p_F + \hat p_F^\dagger + \hat p_B + \hat p_B^\dagger) 
\nonumber \\
&=&- \frac{i}{2a} \left(\begin{array}{ccccccc}
-\lambda & 1 & 0 & \dots & 0 & 0 & 0 \\
-1 & 0 & 1 & \dots & 0 & 0 & 0 \\
 0 &-1 & 0 & \dots & 0 & 0 & 0 \\
 . & . & . & \dots & . & . & . \\
 . & . & . & \dots & . & . & . \\
 0 & 0 & 0 & \dots & 0 & 1 & 0 \\
 0 & 0 & 0 & \dots &-1 & 0 & 1 \\
 0 & 0 & 0 & \dots & 0 &-1 &\lambda_L \end{array}\right) \ , \nonumber \\
i \hat p_I&=&\frac{1}{4}(\hat p_F - \hat p_F^\dagger + \hat p_B - \hat p_B^\dagger)
\nonumber \\
&=&\frac{i}{2a} \mbox{diag}(1,0,0,\dots,0,0,-1) \ .
\label{latticemomentumoperators}
\end{eqnarray}
The resulting momentum operator $\hat p_R + i \hat p_I$ is not Hermitean, but 
has a Hermitean component $\hat p_R$ and an anti-Hermitean component 
$i \hat p_I$, which is diagonal in the position basis. The Hermitean component 
$\hat p_R$ corresponds to a symmetrized forward-backward next-to-nearest 
neighbor derivative that extends over two lattice spacings.

The lattice eigenvalue problem of $\hat p_R$ is given by
\begin{equation}
\hat p_R \phi_{k,x} = \frac{1}{a} \sin(k a) \phi_{k,x} \ ,
\end{equation} 
and the momentum quantization condition takes the form
\begin{equation}
\exp(2 i k L) = \frac{(1 - \lambda \exp(i k a))(1 + \lambda_L \exp(i k a))}
{(\exp(i k a) + \lambda)(\exp(i k a) - \lambda_L)} \ .
\end{equation}
This relation reduces to eq.(\ref{thetaequation}) in the continuum limit.

\end{document}